# Large-bandwidth transduction between an optical single quantum-dot molecule and a superconducting resonator

Yuta Tsuchimoto*, Zhe Sun, Emre Togan, Stefan Fält, Werner Wegscheider, Andreas Wallraff, Klaus Ensslin, Ataç İmamoğlu, and Martin Kroner*

*ETH Zurich, Department of Physics, Zurich, Switzerland*

## Abstract

Quantum transduction between the microwave and optical domains is an outstanding challenge for long-distance quantum networks based on superconducting qubits. For all transducers realized to date, the generally weak light-matter coupling does not allow high transduction efficiency, large bandwidth, and low noise simultaneously. Here we show that a large electric dipole moment of an exciton in an optically active self-assembled quantum dot molecule (QDM) efficiently couples to a microwave resonator field at a single-photon level. This allows for transduction between microwave and optical photons without coherent optical pump fields to enhance the interaction. With an on-chip device, we demonstrate a sizeable single-photon coupling strength of 16 MHz. Thanks to the fast exciton decay rate in the QDM, the transduction bandwidth between an optical and microwave resonator photon reaches several 100s of MHz.

**Introduction**

Because of their bright and fast single-photon emission and the high indistinguishability of emitted photons, self-assembled quantum dots (QDs) have been proposed as an ideal single-photon source for long-distance quantum communication[1]. Harnessing the ability to enhance light-matter interactions in photonic nanostructures, single QDs can serve as both a nearly deterministic single-photon source[2] and an excellent nonlinear element realized by strong coupling to an optical mode[3]. Combined with the recent advances in technologies to tune their resonance frequencies in-situ[4], single QDs are becoming more prominent for constructing quantum networks in the optical domain[1].

For quantum processing at a local node, on the other hand, superconducting qubits[5] and trapped ions[6] are the leading platforms. Particularly, superconducting qubits are attractive because of their high programmability, scalability, fast operational speed, and high fidelity of single- and multi-qubit gates. Circuit quantum electrodynamics in the microwave domain plays a central role in superconducting quantum information processors.

Hybridizing the strengths of both frequency domains – the low attenuation with the enormous bandwidth of optical fibers and the strong computational power of microwave circuits – is a key step towards a distributed quantum information processing network. Many groups have dedicated efforts towards developing various transducers between microwave and optical photons, such as electro-optomechanical systems[7-10], electro-optic systems[11-13], rare-earth ion ensembles[14,15], and magnons[16,17]. Although there are impressive advances such as the conversion of a transmon to an optical photon[7] and low-noise transduction[9], satisfying all the necessary conditions, namely high efficiency, large bandwidth, and low noise, simultaneously is still challenging. The underlying issue is generally weak single-photon coupling strengths.

In this paper, we demonstrate a novel physical platform consisting of an optically active InAs/GaAs QD molecule (QDM) coupled to a superconducting circuit system. The advantages of using an InAs/GaAs QDM for optical-microwave transduction are the following: (i) using a molecular state of the QDs, one can obtain a large electric dipole moment upon single optical photon absorption, which allows for a sizeable coupling strength $g_0$ between a single microwave photon and a single QDM; (ii) the large $g_0$ ensures that additional laser fields are not necessary to effect transduction, which reduces excess noise to a minimum; (iii) the intrinsic bandwidth of the interconnection becomes considerably large (several 100s of MHz) since the exciton recombines on sub-nanosecond timescales; (iv) down-converted single optical photons associated with optical-to-microwave transduction heralds the success of the transduction events, which are essential to realize direct quantum state transfer between distant superconducting qubits with high fidelity[18].

To prove the concept, we fabricate an on-chip device and demonstrate a sizeable coupling strength of 16 MHz as well as transduction between optical and microwave resonator

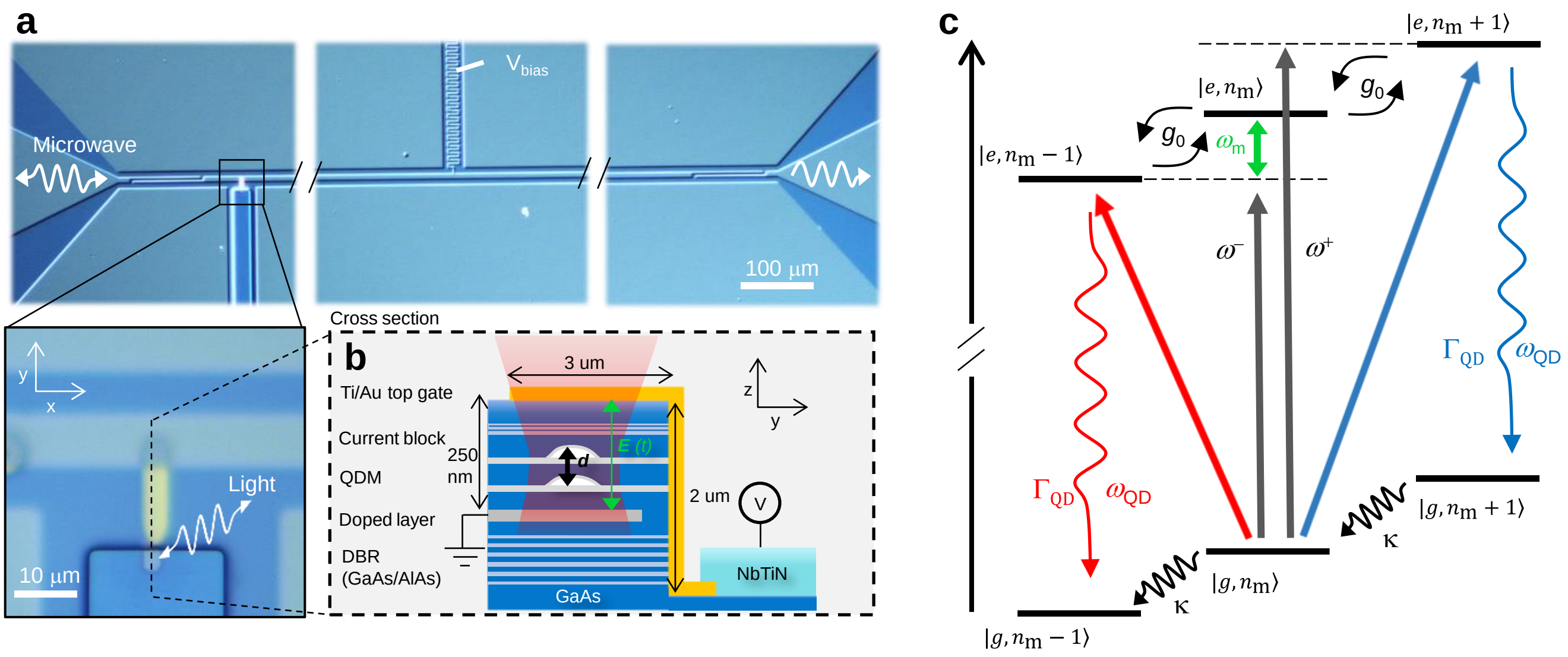

Fig. 1. **a,** Optical microscope images of the transducer system and an enlarged image around QDMs. A λ/2 coplanar waveguide resonator electrically couples to InAs/GaAs QDMs embedded in the mesa structure at the field antinode. **b**, Schematic cross-section (not to scale). The top Ti/Au gate and the bottom doped layer confines a vertical microwave field ***E***, allowing for efficient microwave coupling to the vertical electric dipole moment ***d*** of an optically excited exciton inside the QDM. The bottom DBR mirror, together with the thin Ti/Au top gate, ensures the optical coupling to free space. **c**, Energy diagram of the transduction process. A QDM emits an optical photon upon absorption of a microwave and red-detuned optical photon (microwave to optical transduction highlighted by red) or emits a microwave photon upon absorption of a blue-detuned optical photon (optical to microwave transduction highlighted by blue).

photons with a large bandwidth of several 100s of MHz using a single QDM.

## Device design and transduction scheme

Figure 1a shows our transducer comprising vertically stacked InAs/GaAs QDs and a half-lambda superconducting coplanar resonator (characteristic impedance: $Z_c$ = 130 Ω, resonance frequency: $\omega_m /2\pi$ = 9.7 GHz, total and external decay rates: $\kappa/2\pi$ = 138 MHz and $\kappa_{ext}/2\pi$ = 8.3 MHz) on GaAs. The QDs are embedded in the mesa-like heterostructures located at an electric-field antinode nearby the central line of the resonator. On the top surface of the mesa structure, a 15 nm-thick Ti/Au layer (top gate) was deposited, which is connected to the central line of the resonator at the field antinode and feeds microwave photons to the QDs from the resonator (see the enlarged image). Below the QDs, a doped n-type GaAs layer (back contact) is connected to an electrical ground (see Fig. 1b). The overlap of the top gate and the back contact generates a local vertical microwave field at the QD position. We apply a bias voltage to the QDs through the bias line located at the node of the resonator in order to tune the electronic levels of top and bottom QDs into resonance to form a QDM. A vertically oriented permanent electric dipole of an indirect exciton in the QDM interacts with the vertical microwave field efficiently via the DC-Stark effect (see Fig. 1b). A distributed Bragg reflective (DBR) mirror located below the back contact, together with the thin top gate, forms an optical cavity that efficiently couples the QD emission to free space. Such low-Q optical cavities with a leaky top mirror have been successfully used for distant QD spin entanglement[19] as well as absorption of a photonic qubit by a single QD[20]. We remark that our device structure in Fig. 1a resembles a capacitively coupled microwave qubit-resonator system[21-23]. Thus, one could incorporate a superconducting qubit into the other antinode of our resonator and perform qubit-to-optical photon transduction via the resonator bath.

Figure 1c shows the energy diagram of the transduction scheme. $|g\rangle$ denotes the ground state of the QDM, and $|n_m\rangle$ represents the state of the microwave resonator with $n_m$ photons. The QDM has symmetric $|f\rangle$ (not shown) and anti-symmetric $|e\rangle$ excited molecular states because of the electron tunnel coupling between the QDs. Ideally, the frequency difference $\Delta_{ef}$ between $|e\rangle$ and $|f\rangle$ should match the microwave resonator frequency $\omega_m$, as proposed in Ref. 18. In this case, it should be possible to achieve the transduction with unity internal efficiency[18] and demonstrate optical sideband cooling of the microwave resonator. Here we present a proof-of-principle experiment where, for the ease of fabrication, we omitted the condition of the resonance between $\Delta_{ef}$ and $\omega_m$.

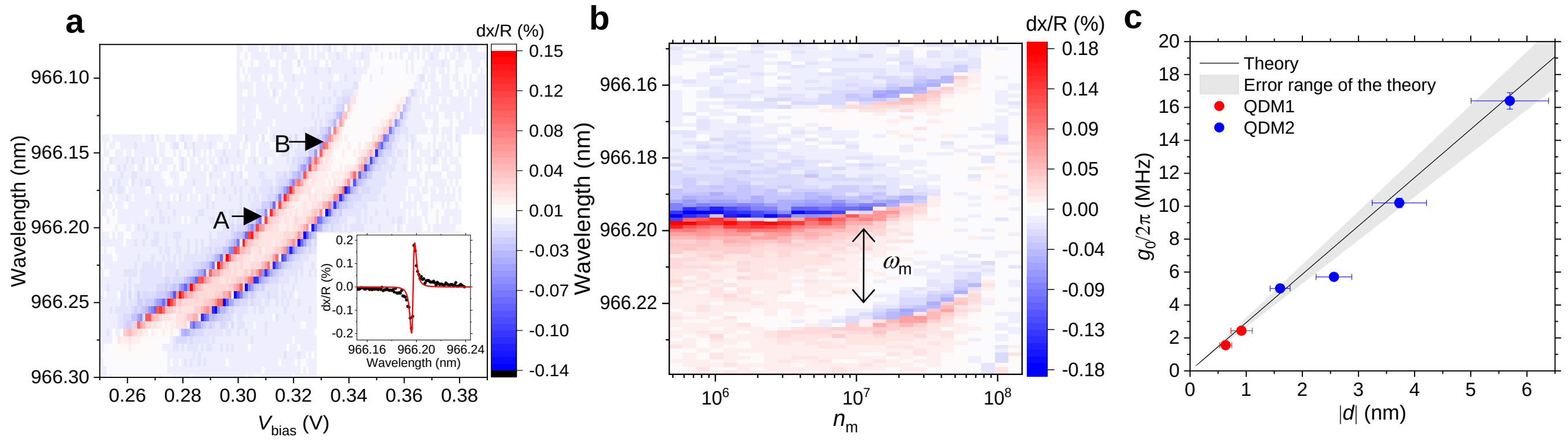

Fig. 2. **a,** Bias voltage ($V_{bias}$) dependence of DR for a QDM measured by lock-in detection. **b**, DR signal measured at the point A in **a** while driving the QDM with increasing microwave resonator photon number ($n_m$). **c**, Relationship between the estimated single-photon coupling strengths ($g_0$) and the dipole sizes ($|d|$) for QDM1 (red dots) and QDM2 (blue dots) in the same device. The error bars represent the standard errors of the fits used for the estimation. The solid black line and shaded area show the theoretically estimated $g_0$ and its systematic error range.

We used single QDMs with a significant frequency separation $\Delta_{ef} \sim 2\pi \times 500$ GHz $\gg \omega_m$. We drive the system by a weak laser with a frequency close to the |g> → |e> transition frequency. In this case, the system is approximated by an effective two-level system modulated by the microwave field. The Hamiltonian for the system reads

$$\hat{H} = \hbar\omega_{QD}\hat{\sigma}_{ee} + \hbar\omega_L \hat{a}^\dagger\hat{a} + \hbar\omega_m \hat{b}^\dagger\hat{b} + \hbar g_0 \hat{\sigma}_{ee}(\hat{b}^\dagger + \hat{b}) + \hbar\int d\omega\, \kappa_{QD}(\omega)[\hat{\sigma}_{ge}\hat{a}^\dagger + \hat{\sigma}_{eg}\hat{a}], \quad (1)$$

where $\omega_{QD}$ and $\omega_L$ are the resonance frequency of a QDM in state |e> and the laser frequency, respectively. $\sigma_{ee}$ represents the projection operator to state |e>, and $\sigma_{ge}$ and $\sigma_{eg}$ are lowering and raising operators between |*g*> and |*e*>. $a^\dagger$ ($b^\dagger$) and $a$ ($b$) are creation and annihilation operators for optical (microwave) photons. The fourth term describes the event where |*e*> virtually absorbs or emits a microwave photon at a single-photon coupling strength $g_0 = e\boldsymbol{d} \cdot \boldsymbol{E}_{mw}$ where $e$, $\boldsymbol{d}$, and $\boldsymbol{E}_{mw}$ are the elementary charge, the size of the dipole moment of the QDM, and the vacuum fluctuation of the microwave field at the QDM. The last term represents the coupling of the QDM to an optical field[24] and $\kappa_{QD}(\omega) \approx \sqrt{\Gamma_{QD}/2\pi}$ under the standard Markov approximation, where $\Gamma_{QD}$ is the decay rate of the exciton in the QDM. For transduction, we drive the QDM with blue- or red-detuned optical photons at frequencies $\omega^{\pm} = \omega_{QD} \pm \omega_m$. The QDM converts optical (microwave) photons to microwave (optical) photons upon absorbing the blue-(red-)detuned photons.

**Microwave modulation of an exciton**

First, we demonstrate the coupling between an exciton and the superconducting resonator in order to estimate $g_0$. In the following experiments, we keep the optical laser power within the linear regime of QD absorption. Figure 2a shows the bias voltage ($V_{bias}$) dependence of the differential reflection (DR) signal of a negatively charged exciton in a QDM (QDM1). A clear dispersive resonance due to the interference between the laser reflection and the QD emission is visible. The two bright curves separated by 0.02 V are replicas of the resonance due to the kHz modulation of $V_{bias}$ for the lock-in measurement of the DR signal (see Method for the details). The excitonic resonance shows a more significant DC-Stark shift with increasing $V_{bias}$ because the electron levels of the two QDs come close to resonance, increasing the indirect exciton component. From the DC-Stark shift $\delta\epsilon_{Stark} = e\boldsymbol{d} \cdot \boldsymbol{E}_{bias}$, we can estimate the effective dipole size. QDM1 shown in Fig. 2a exhibits $|d|$ ~ 0.9 nm at point B (wavelength: 966.13 nm). In another molecule, QDM2 (shown in Supplementary Fig. S2), we find $|d|$ ~ 5 nm, which is several orders of magnitude larger than the permanent electric dipole moment of typical single QDs. In principle, the maximum dipole size should be up to ~ 12 nm (corresponding to the separation length of the two dots in this device, see Supplementary Fig. S1b) by increasing $V_{bias}$. The maximum bias that can be applied and thereby the maximum dipole size is, however, limited by the charge stability in the QDM due to tunnel coupling to the electron reservoir. The inset shows a wavelength scan at a fixed $V_{bias}$. The linewidth $\Gamma_{QD}^{inh}/2\pi$ obtained by fitting a dispersive Lorentzian to the data is 720 ± 30 MHz which is about twice the transform-limited linewidth that one would obtain if the QDM line broadening were exclusively due to radiative decay[25,26]. This inhomogeneous broadening is because of the well-known low-frequency noise in the environment of the QDM due to charge fluctuations[26]. In principle, it can be eliminated by sample optimization[25,27]. We note that the inhomogeneous broadening observed here is not due to the device fabrication processes but rather the nature of the as-grown QDs themselves[25], which we confirmed by characterizing the QDs on an unprocessed piece of the wafer.

To demonstrate the coupling between the exciton and the superconducting resonator, we investigated the exciton spectrum around $\omega_{\mathrm{QD}}$ by monitoring the DR signal while resonantly driving the superconducting resonator at the frequency $\omega_{\mathrm{m}}$. The coherent oscillation of the microwave field at the QDM changes the exciton frequency via the Stark shift as $\widetilde{\omega}_{\mathrm{QD}} = \omega_{\mathrm{QD}} + 2g_0\sqrt{n_{\mathrm{m}}}\sin(\omega_{\mathrm{m}}t)$ . This frequency modulation of the exciton leads to the appearance of sidebands in the absorption spectrum. In Figure 2b, we plot the DR signal taken at point A from Fig. 2a, as a function of the microwave mean photon number $n_{\mathrm{m}}$ in the resonator (see Supplementary IV-B for the calibration of $n_{\mathrm{m}}$). The sidebands start to appear at $n_{\mathrm{m}} \sim 3 \times 10^6$. The amplitudes of the resonance and the sidebands oscillate proportionally to the Bessel function of the first kind $J_l(2g_0\sqrt{n_{\mathrm{m}}}/\omega_{\mathrm{m}})$. Here, $l$ = 0 and 1 correspond to the resonance and its first sidebands. With this, we can fit the obtained amplitudes as a function of $n_{\mathrm{m}}$ and estimate $g_0$ (see Supplementary Fig. S6). Figure 2c shows the estimated $g_0$ for various dipole sizes obtained from the two different QDMs in the same device. The red dots correspond to points A and B from QDM1 in Fig. 2a, and the blue dots are from QDM2 shown in Supplementary Section II. The error bars on $g_0$ and $|\boldsymbol{d}|$ are standard errors of the fits. The solid black line shows a theoretically estimated coupling strength (see Supplementary Section V), and the grey shadow area around the solid line displays an estimated systematic error range. We show more details of the theory and the error calculations in Supplementary Section II and V. The experimentally determined $g_0$ agrees very well with the theoretical prediction. The obtained $g_0$ increases from 1.5 to 16 MHz proportionally to the dipole size. The large single-photon coupling strength obtained here is on the same order as those realized by collective excitations of spins (magnon systems)[16,17] and several orders of magnitude larger than those of other transducers.

**Microwave-optical transduction**

We define the transduction process as the conversion of an optical photon to a microwave photon in the resonator or conversely, the conversion of a microwave photon in the resonator to an optical photon upon arrival of a red-detuned optical photon. We do not consider injecting or extracting microwave photons from the resonator through the finger capacitors since we envision transferring a microwave resonator photon between a QDM and a superconducting qubit coupled in the same resonator but at different antinodes. Theoretically, the internal efficiency $\eta_{\mathrm{int}}$ of that transduction process can be calculated from Eq. (1) (see Supplementary Section VI-A) and is given by

$$\eta_{\mathrm{int}} = \frac{4C_0}{(1 + C_0)^2 + 4\left(\frac{\omega_{\mathrm{m}}}{\Gamma_{\mathrm{QD}}}\right)^2}, \tag{2}$$

where $C_0 = 4{g_0}^2/(\Gamma_{\mathrm{QD}} + \kappa)\Gamma_{\mathrm{QD}}$ is the single-photon cooperativity between the QDM and the superconducting resonator. The second term in the denominator of Eq. (2) appears because the QDM absorbs the detuned incoming optical photons off-resonantly via the sidebands of the |g> → |e> transition when $\Delta_{\mathrm{ef}} >> \omega_{\mathrm{m}}$. With the assumption of $\omega_{\mathrm{m}}/\Gamma_{\mathrm{QD}} \gg 1 + C_0$, Eq. (2) simplifies to

$$\eta_{\mathrm{int}} = C_0\left(\frac{\Gamma_{\mathrm{QD}}}{\omega_{\mathrm{m}}}\right)^2. \tag{3}$$

If an exciton in the QDM is coherently driven by microwave or optical photons, the coupling strength increases by the $\chi^{(2)}$ type interaction (see Supplementary Section VI). The multiphoton cooperativity is then given by $C = (n_{\mathrm{m}} + k)n_{\mathrm{o}}C_0$, where $n_{\mathrm{o}}$ is the optical photon number coupled to the QDM in the leaky optical cavity. $k$ = 0 and 1 correspond to the microwave-to-optical and optical-to-microwave transductions, respectively. Then the internal transduction gain can be written as $G_{\mathrm{int}} = (n_{\mathrm{m}} + k)n_{\mathrm{o}}\eta_{\mathrm{int}} = C(\Gamma_{\mathrm{QD}}/\omega_{\mathrm{m}})^2$ when $\omega_{\mathrm{m}}/\Gamma_{\mathrm{QD}} \gg 1 + C$, predicting a linear increase of $G_{\mathrm{int}}$ with $n_{\mathrm{m}}$.

For the microwave-optical transduction experiment, we tuned the laser frequency to the red- or blue-sideband ($\omega^{\pm} = \omega_{\mathrm{QD}} \pm \omega_{\mathrm{m}}$) of the excitonic resonance in QDM1 or QDM2. In order to investigate the transduction mediated by the single optical photon interaction, we kept $n_{\mathrm{o}} \sim 1$ throughout the measurements by keeping the laser power within the linear response regime of the QDM (≤ 30 nW). (see Supplementary Section VI-B for the results with larger $n_{\mathrm{o}}$). On the other hand, we drove the QDM with large $n_{\mathrm{m}}$ to obtain $G_{\mathrm{int}}$ in a detectable range. Experimentally, we estimated $G_{\mathrm{int}}$ by measuring the up- or down-converted optical photon rate (red or blue wavy arrows in Fig. 1b) since those individual single photons tell us the success of every transduction event upon detection (see Method for the measurement setup). We normalized the count rate of the converted photons by the independently detected photon rate from the resonantly excited |e> → |g> transition without microwave drive. This resonant emission rate used for the normalization corresponds to the detectable converted photon rate at $\eta_{\mathrm{int}} = 1$.

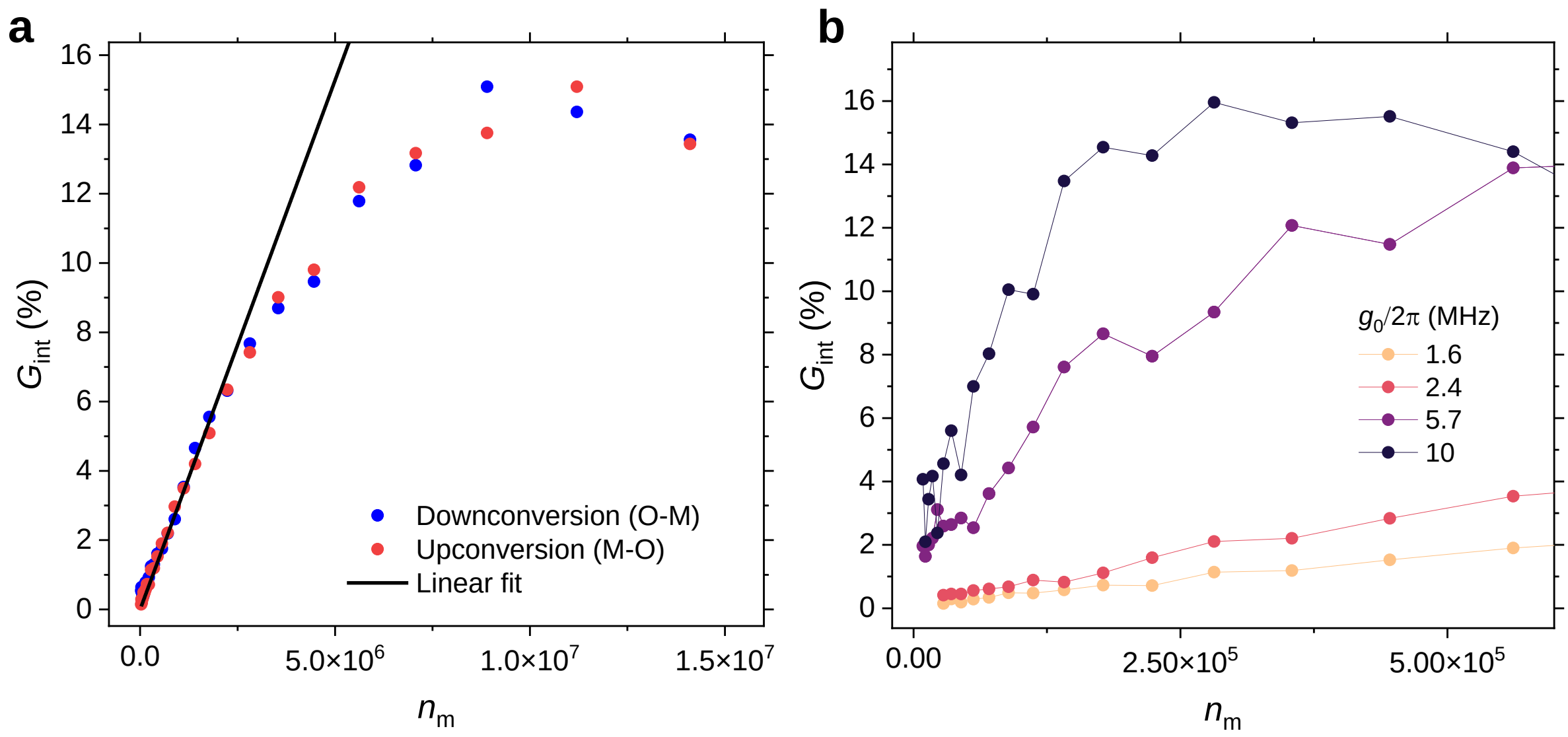

Fig. 3. **a,** Microwave mean photon number ($n_m$) dependence of the internal transduction gain ($G_{int}$) for up- and down-conversion with the red- or blue-detuned optical drives, corresponding to the microwave-to-optical (M-O) or optical-to-microwave (O-M) transduction gain, respectively. The solid black line exhibits the fit for the linear regime. **b**, single-photon coupling strength ($g_0$) dependence of $G_{int}$ for the M-O transduction measured with QDM1 and QDM2 at various $n_m$.

Figure 3a shows $G_{int}$ as a function of $n_m$ at $g_0/2\pi = 1.6$ MHz. For both red- and blue-detuned cases, $G_{int}$ increases linearly following Eq. (3) in the range of $1 + C \ll \omega_m/\Gamma_{QD}$ as shown by the linear fit (solid black line) and deviates from it when $n_m \gtrsim 2 \times 10^6$ due to the higher-order scattering processes. We estimated $\eta_{int} = 3 \times 10^{-8}$ by using the relation $\eta_{int} = \lim_{n_m \to 0 \text{ or } 1} G_{int}$ in the linear regime for both blue- and red-detuned cases. We remark that the estimated $\eta_{int}$ is the value relying only on single optical and microwave photon interactions and is not enhanced by coherent optical or microwave pump photons, which are detrimental to low-noise operation. We find optical outcoupling efficiency $\eta_o = 2 \times 10^{-3}$ by measuring the input optical power and the resonantly emitted photon counts from QDM1, which leads to the efficiency of $\eta = \eta_o \eta_{int} = 6 \times 10^{-11}$. We enhance $\eta_{int}$ by increasing $g_0$ as shown in Fig 3b. $G_{int}$ becomes larger in the linear regime with increasing $g_0$, thereby, $\eta_{int} = \lim_{n_m \to 0 \text{ or } 1} G_{int}$ goes up to ~ $10^{-6}$ at $g_0/2\pi = 10$ MHz for both blue- and red-detuned cases (the blue-detuned case not shown here). $\eta_o$ decreases gradually with increasing indirect exciton component because of the larger exciton linewidth arising from environmental charge noise. (See Supplementary Section II).

## Transduction bandwidth

The bandwidth of the transduction process – an absorbed optical photon is converted to a microwave photon in the coplanar resonator and vice versa – is reflected in the bandwidth of the optical excitation of the QDM since the end of the transduction process is associated with the exciton recombination[18]. We measured the transduction bandwidth by counting the up- (top panel in Fig. 4a) and down-converted photons (bottom panel) as a function of the laser frequency around the red- and blue-detuned sidebands ($\omega_L = \omega^{\pm} + \Delta\omega$) at $g_0/2\pi = 1.6$ MHz. We fixed the microwave drive frequency at the resonance and the photon number inside the resonator to $n_m = 1.2 \times 10^7$. The emission spectra were measured through an etalon in order to filter the elastically scattered photons. The obtained spectra through the etalon were normalized by the etalon's transmission spectrum (linewidth: 1.7 GHz, see Method for the measurement setup). From the Lorentzian fit to the normalized spectra shown in Fig. 4a, we estimated the linewidth to be FWHM = 840 ± 90 MHz for the upconversion and FWHM = 770 ± 160 MHz for the downconversion, which agree well with the expected linewidth $(\Gamma_{QD}^{inh} + \kappa)/2\pi$ ~ 860 ± 30 MHz. The measured FWHM here is the upper limit of the transduction bandwidth $B/2\pi = (\Gamma_{QD} + \kappa)/2\pi$ since $\Gamma_{QD}^{inh}$ includes the low-frequency noise component, as mentioned before. The presence of the low-frequency noise component reduces the optical-microwave transduction efficiency, and the presence of the microwave photon decay deteriorates state transfer efficiency and fidelity between a microwave photon temporally stored in the resonator and a superconducting qubit which would be coupled to the other antinode of the resonator. For optimum efficiency and fidelity, one must operate with a transform-limited QDM exciton spectrum[25] and a low-loss superconducting resonator. Typically, the pure decay components of self-assembled InAs/GaAs QDs in similar heterostructures is $\Gamma_{QD}/2\pi$ ~ 300 MHz[25,28]. In this case, the bandwidth of $B/2\pi = \Gamma_{QD}/2\pi$ ~ 300 MHz is smaller than the measured FWHM but still sufficiently large. $B/2\pi$ ~ 300 MHz is enough to perform the transduction before the lifetime of a superconducting qubit as well as to incorporate time and frequency multiplexing techniques. This intrinsically large bandwidth is an advantage of our system. Further improvement of $B$ can be achieved by exploiting the Purcell effect by incorporating the QDM into a high-Q optical cavity.

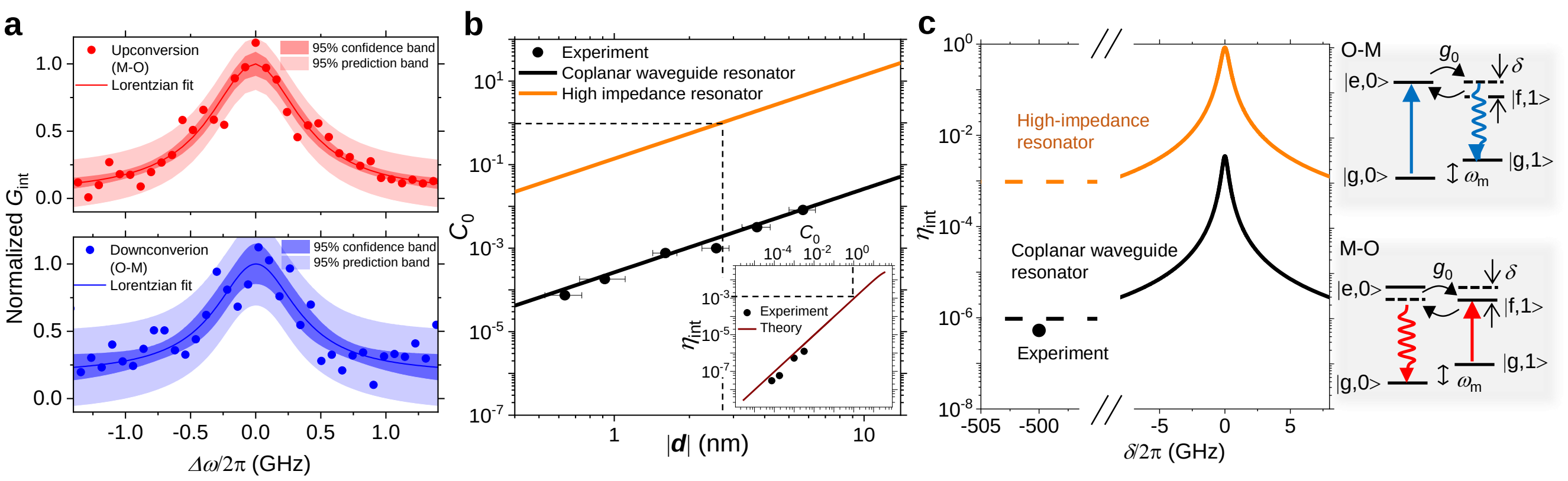

Fig. 4. **a**, Normalized internal transduction gain $G_{int}$ for up-(top panel) and down-conversion (bottom panel) measured as a function of the red- or blue-detuned laser frequency ($\omega_L = \omega^{\pm} + \Delta\omega$) at a fixed microwave drive frequency, power, and bias voltage. The data was normalized by the Lorentzian fits. The up- and down-conversion correspond to the microwave-to-optical (M-O) and optical-to-microwave (O-M) transduction, respectively. The dark and light red (blue) shaded areas show the 95 % confidence and prediction bands of the fit shown as the solid red (blue) line. The signal to noise ratio of the down-converted photons is worse than that of the up-converted photons simply because the excitation laser power (~ 3 nW) used particularly for this down-conversion experiment is an order of magnitude smaller. **b**, Dipole size (|$d$|) dependences of the single-photon cooperativity ($C_0$). We calculated the solid black line and points using the theoretically and experimentally estimated $g_0$ shown in Fig. 2c, $\Gamma_{QD}/2\pi$ = 300 MHz, and $\kappa/2\pi$ = 138 MHz. The solid orange line shows the dependence when a QDM couples to a high-impedance superconducting resonator with $\kappa << \Gamma_{QD}$. The dashed line indicates $C_0$ = 1 at |***d***| = 2.6 nm. The inset shows $\eta_{int}$ as a function of $C_0$. The solid brown line is the theoretically predicted result using Eq. (2) and $C_0$ shown in the main panel, and the black dots represent the experimentally estimated $\eta_{int}$ from Fig. 3b. The dashed line indicates $\eta_{int}$ ~ $10^{-3}$ at $C_0$ = 1. **c**, Theoretically predicted detuning ($\delta = \omega_m - \Delta_{ef}$) dependence of $\eta_{int}$ at |***d***| = 2.6 nm. The dashed lines are far-detuned cases (|$\delta$| >> $\omega_m$) for the current device (black) and a QDM with a high-impedance resonator (orange). The solid lines are close-to-resonance cases (|$\delta$| < $\omega_m$) where state |f> plays a role. The black dot indicates the experimentally estimated $\eta_{int}$ with the current device at |***d***| = 2.6 nm. The energy diagrams next to the graph show the M-O (red) and O-M (blue) transduction processes with the detuning |$\delta$| < $\omega_m$.

## Prospects

While the transduction bandwidth is orders of magnitude larger than other systems, the transduction efficiency demonstrated here has much room to improve. We emphasize that the experimentally demonstrated efficiency is the result of the single optical and microwave photon interactions. Here we outline that one can improve the efficiency to close-to-unity without using additional coherent pump fields in contrast to typical transducers. The absence of additional pump fields should eliminate transduction noise arising from optical heating and higher-order scattering processes.

First, one can improve the efficiency by enhancing $g_0$ further. The data points in Fig. 4b shows the single-photon cooperativity as a function of dipole size |***d***|, obtained from the measured $g_0$ by $C_0 = 4g_0^2/(\Gamma_{QD} + \kappa)\Gamma_{QD}$. Here we employ the transform-limited linewidth $\Gamma_{QD}$ rather than $\Gamma_{QD}^{inh}$ and include the contribution of the low-frequency noise into the optical outcoupling efficiency $\eta_0$. The solid black line is calculated in the same way for the theoretically estimated $g_0$ shown in Fig. 2c. The demonstrated $C_0$ in this paper ranges from ~ $10^{-4}$ to ~ $10^{-2}$ for |***d***| ~ 0.6 to ~ 5 nm. The corresponding $\eta_{int}$ increases from ~ $10^{-8}$ to $10^{-6}$, as displayed in the inset of Fig. 4b. To achieve a larger $C_0$, one can integrate the QDM into a high-impedance superconducting resonator[29,30] with a negligible $\kappa$. For example, if $Z_c$ = 5000 Ω and the electric field at a QDM is confined to ~ 80 nm, $g_0$ increases by a factor of ~ 20. $C_0$ for the high-impedance device follows the solid orange line in Fig 4b, and we find $C_0$ = 1 at |***d***| ~ 2.6 nm. In this case, the

corresponding $\eta_{\rm int}$ calculated by Eq. (2) reaches ~ $10^{-3}$. We have fabricated such a device including gate-tunable QDs and have observed negligibly small loss $\kappa \ll \Gamma_{\rm QD}$ (this is work in preparation).

The internal efficiency $\eta_{\rm int}$ can be further improved by decreasing $\Delta_{\rm ef}$. Figure 4c shows the theoretically expected $\eta_{\rm int}$ as a function of the detuning $\delta = \omega_{\rm m} - \Delta_{\rm ef}$. The black and orange lines indicate the cases for the current device and a device with the high-impedance resonator for $|\boldsymbol{d}|$ = 2.6 nm, respectively. When $|\delta| >> \omega_{\rm m}$, the transduction process relies on the single-optical resonant process. Here, an incoming optical photon is absorbed off-resonantly via its red- or blue-sideband of the |g> → |e> transition, whereas a converted optical photon is emitted resonantly through the transition (see Fig. 1c for the energy diagram). The efficiency is expressed by Eq. (2) and does not depend on the detuning (as indicated by the dashed lines in Fig. 4c). Only when $|\delta| < \omega_{\rm m}$, the efficiency is enhanced significantly because of the double-optical resonant process: incoming and converted optical photons interact with the |g> → |e> and |g> → |f> transitions resonantly (see solid lines. Note that the calculation is accurate only within the rotating wave approximation, $|\Delta_{\rm ef} - \omega_{\rm m}| << \Delta_{\rm ef} + \omega_{\rm m}$. see Supplementary IV). At the resonance condition of $\delta = 0$, the internal efficiency is given by $\eta_{\rm int} = 4C_0/(1+C_0)^2$, which leads to $\eta_{\rm int} = 1$ for $C_0 = 1$ because of quantum interference as studied in Ref. 18. Comparing it with the far-detuned case described by Eq. (2), the enhancement factor for the efficiency is $(\omega_{\rm m}/\Gamma_{\rm QD})^2 \sim 10^3$. To experimentally satisfy the resonant condition of $\delta = 0$, one can reduce $\Delta_{\rm ef}$ by increasing the tunnel barrier between the QDs by inserting an Al-rich layer between each QD layer, as shown in Supplementary Section III. The deterministic positioning technique[27] of QDs enables integrating a particular QDM with the right frequency $\Delta_{\rm ef} = \omega_{\rm m}$ into a superconducting resonator. In short, $\eta_{\rm int}$ reaches unity by combining the right QDM with the high-impedance resonator that allows achieving $C_0 = 1$ (solid orange line).

Finally, improving $\eta_{\rm o}$, which accounts for optical frequency mismatch due to the inhomogeneous broadening and spatial mode mismatch, is also an important aspect. One can reduce the low-frequency noise from charge fluctuations by introducing a careful surface treatment[31] and shielding by metallic gates[27]. We have previously achieved superior spatial mode matching by simply using an optimized DBR mirror at the bottom of QDMs, and recorded $\eta_{\rm o}$ of up to 16 %[20]. Using a photonic waveguide, a micropillar, or an open-access microcavity is also a promising route, and an end-to-end efficiency of 57 % has been reported recently[32].

For actual quantum information transfer, one can employ heralding photons associated with the optical-to-microwave transduction to improve the transfer fidelity. For transduction that relies on an optical pump field to achieve high transduction efficiency, the heralding optical photon is emitted into the coherent optical pump field; as such, the information carried by this photon is lost. In the case of a QDM, which does not require additional pump fields, one can directly transfer a quantum state between distant superconducting qubits through an optical photon by detecting heralding events. Specifically, we consider the case where a QDM and a superconducting qubit are coupled to different antinodes of the same low-loss coplanar waveguide resonator ($\kappa << \Gamma_{\rm QD}$). In the optical-to-microwave state transfer, one could employ a time-bin optical qubit which would have two time windows separated by a time delay $\tau$ that is, at least, longer than the QDM lifetime. After the first time-bin optical photon is converted to a microwave resonator photon, the resonator population is transferred to an excited state of the superconducting qubit by SWAP operation. Subsequently, the second time-bin optical photon is transferred to another excited state of the superconducting qubit. The success events of the transduction from an optical photon to a microwave photon can be heralded by detecting a down-converted optical photon after erasing which-time information with an unbalanced Mach-Zehnder interferometer. In the reverse process, quantum information in a superconducting qubit is converted into a microwave time-bin qubit. While the details of the process would rely on the specific features of the superconducting qubit, we could consider a protocol where a single microwave photon is generated at time $t_1$ if the qubit is initially in state |1>. This microwave photon is then converted into an optical photon using a laser-pulse incident at time $t_1 + T$; here, it is essential that $T << \kappa^{-1}$. At time $t_2 > t_1 + T$, a second microwave photon is generated if the superconducting qubit is in state |0>. Finally, upon application of a second laser pulse at time $t_2 + T$, the quantum information initially stored in the superconducting qubit is fully converted into an optical time-bin qubit. Since each laser pulse is long compared to the QDM lifetime but shorter than each time window, the precise timing of the optical pulses is not necessary. Here, in the reverse process, the heralding photons do not exist. However, when an upconverted optical photon is eventually downconverted in the other transducer coupled with a superconducting resonator and qubit, the success of both up and downconversions are heralded. In this scheme, one must discard 50 % of photons as a cost of the probabilistic erasure step, but the transfer fidelity becomes high. We note that the narrow linewidth of a low-loss superconducting resonator does not limit the state transfer speed but rather the anharmonicity of the superconducting qubit sets the upper limit of the speed (several 100s of MHz for transmons[33]) in order to avoid higher-order excitations of the qubit (see the theoretical study for more details[18]).

One could also use a QDM as a fast single microwave photon source driven by an optical laser. Thanks to the low thermal conductivity of optical fibers and the low laser power required for the transduction, optically driven single microwave photon sources could have an advantage of reduced heat load compared with the ones driven by microwaves. To solve the heat load problem on a large-scale quantum computer, some groups have investigated optical readout and control schemes of superconducting qubits recently[34,35].

## Conclusion

We demonstrated a device that features a considerable single-photon coupling strength of up to 16 MHz and a transduction bandwidth of several 100s of MHz, which are few orders of magnitude larger than other systems. To this end, we fabricated an on-chip transducer by integrating self-assembled QDMs on a superconducting coplanar waveguide resonator. By combining the above-mentioned improvements, one can achieve a conversion efficiency of several 10s of % between an optical and a microwave photon without additional coherent pump fields. We stress that these improvements have already been demonstrated individually. The remaining challenge is to design and integrate all in one device on a chip. The results obtained here open a way towards a large-bandwidth and low-noise interconnection between the microwave and optical domains and may even facilitate other applications of QDs in the microwave domain, such as optically driven fast single microwave photon sources.

## Methods

### Measurement setup

Figure 5 shows a schematic of our experimental setup. A microwave drive from a signal generator is fed to the device through a high pass filter and coaxial cables. The transmission from the device is amplified at room temperature and fed back to a signal analyzer. Optical lasers were coupled to the QDMs through an objective lens (Numerical aperture: 0.68). We focused the laser on a QDM under the top gate by moving the device with piezo stages while imaging the surface with a light-emitting diode (LED) and an imaging charge-coupled device (CCD) camera. The QD emission and the laser reflection background were collected by the objective and sent to one of the detection setups. In order to find QDs under the top gate, we performed photoluminescence measurement with the 780 nm laser while scanning the piezo stages. The laser reflection was removed by a long-pass filter, and only the emission from QDs reached a grating spectrometer and a charge-coupled device camera. For the DR measurements, an exciton in QDM1 or QDM2 was excited by a resonant laser. The resonantly excited exciton was modulated by a square voltage wave (amplitude: 0.2 V, frequency: 3.977 kHz) through the bias line. The modulated emission and the laser background was detected with a photodiode followed by an amplifier. A lock-in amplifier demodulated the signal and extracted it from the laser background. In the transduction experiment, an exciton in QDM1 or QDM2 was excited by a red- or blue-detuned laser ( $\omega_L = \omega_{QD} \pm \omega_m$ ) with the microwave drive on. The laser reflection background was suppressed by a factor of ~ $10^6$ using cross-polarized excitation and detection. The converted single photons at $\omega_{QD}$ and remaining laser reflection background passed through an etalon (linewidth: 1.7 GHz, free spectral range (FSR): 34 GHz) and were detected by an avalanche photodiode (APD). We used an etalon with a relatively wide transmission window and FSR so that the whole spectral range of the QD emission is transmitted while the elastically scattered photons are suppressed. An APD detected converted photons together with the residual laser reflection background from the sample surface and the dark counts of the APD. From this signal, we subtracted the dark counts and the laser background in a post-processing step to obtain the up or downconverted photon counts. This background signal was obtained by gate voltage tuning of the exciton sideband frequency out of resonance with the laser frequency ( $\omega_{QD} \pm \omega_m \neq \omega_L$ ). In order to estimate $G_{int}$, we normalized the count rate of the converted photons by the count rate obtained by resonantly exciting the |e> -|g> transition of the QDM through the same setup. Here, the laser frequency was shifted from the sidebands to the QD resonance frequency $\omega_{QD}$ and the microwave drive was switched off. This resonant emission tells us the maximum photon rate that we can detect with the setup, which corresponds to the detectable converted photon rate at $\eta_{int} = 1$. Since the setups for the resonant emission measurement and the converted photon measurement are identical, the imperfections of the outcoupling and the setup efficiency were reliably eliminated in $G_{int}$.

## Data availability

The datasets generated during the current study are available at DOI: xxx

## Acknowledgements

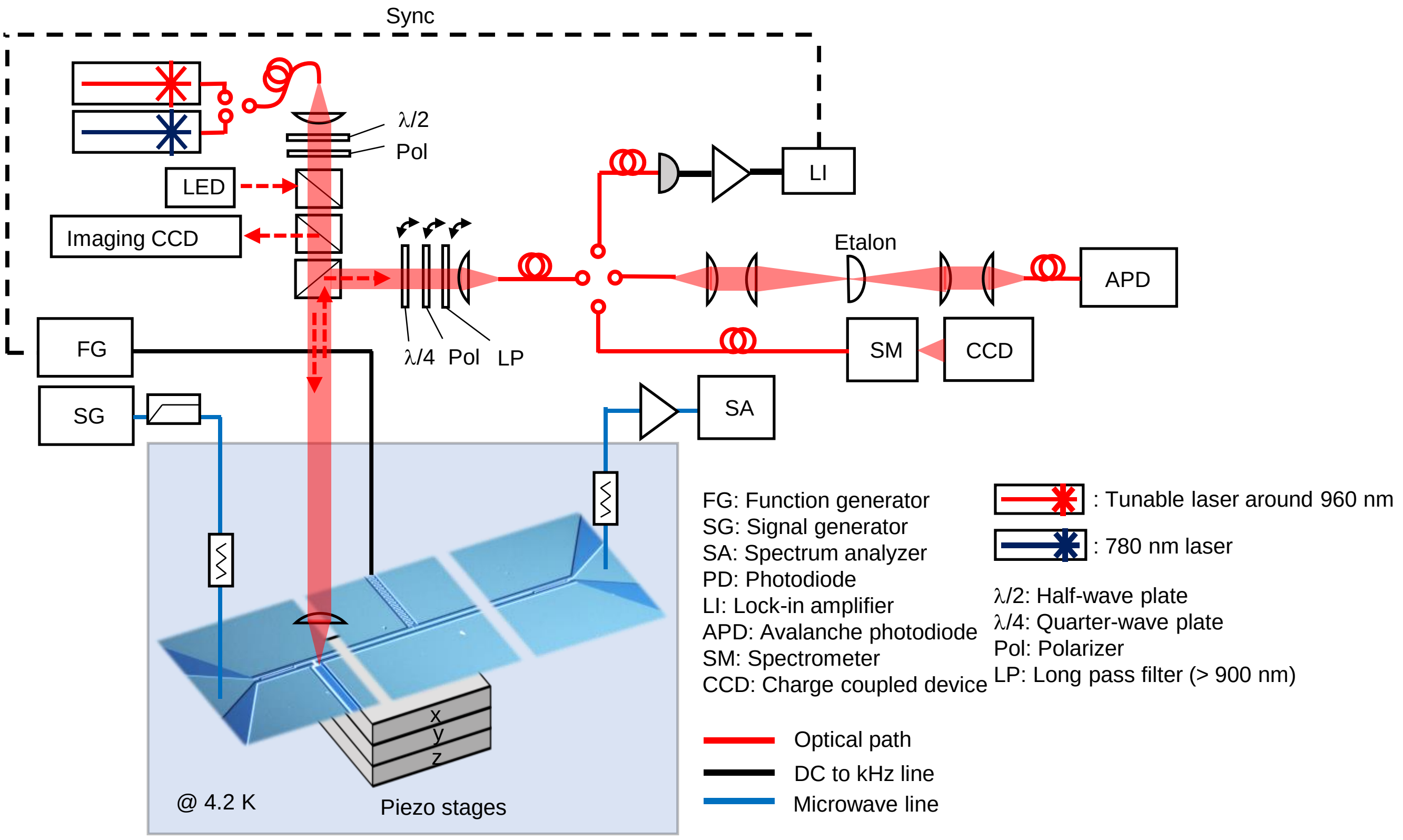

Fig. 5. Schematic of the measurement setup. The sample is located at 4.2 K in a bath cryostat. The system comprises the setups for microwave drive and detection, photoluminescence measurement, DR measurement, and transduction experiment.

This publication was produced within the scope of the National Research Program NCCR QSIT, which was funded by the Swiss National Science Foundation. We would like to thank the Swiss National Science Foundation for their financial support. We acknowledge helpful discussion on superconducting circuits with Pasquale Scarlino and Andreas Landig and thank Xiaofu Zhang for superconducting film characterization at an early stage of the project. We appreciate discussions with Patrick Knüppel and his early-stage contribution to QD characterization. We thank the Cleanroom Operations Team of the Binnig and Rohrer Nanotechnology Center (BRNC) and FIRST laboratory for their help and support.

**Contributions**

Y.T. performed the sample fabrication, the measurements, and the analyses. Z.S. contributed to experiments at an early stage of the project. E.T. supported the sample fabrication. S.F. grew quantum dots under the supervision of W.W; A.W. and K.E. supported the fabrication of the superconducting circuit. Y.T. and M.K. wrote the manuscript. Y.T., A.I., and M.K. supervised the project.

## References

1 Lodahl, P. Quantum-dot based photonic quantum networks. *Quantum Science and Technology* **3**, doi:10.1088/2058-9565/aa91bb (2018).

2 Arcari, M. *et al.* Near-unity coupling efficiency of a quantum emitter to a photonic crystal waveguide. *Phys Rev Lett* **113**, 093603, doi:10.1103/PhysRevLett.113.093603 (2014).

3 Najer, D. *et al.* A gated quantum dot strongly coupled to an optical microcavity. *Nature* **575**, 622-627, doi:10.1038/s41586-019-1709-y (2019).

4 Grim, J. Q. *et al.* Scalable in operando strain tuning in nanophotonic waveguides enabling three-quantum-dot superradiance. *Nat Mater* **18**, 963-969, doi:10.1038/s41563-019-0418-0 (2019).

5 Arute, F. *et al.* Quantum supremacy using a programmable superconducting processor. *Nature* **574**, 505-510, doi:10.1038/s41586-019-1666-5 (2019).

6 Wright, K. *et al.* Benchmarking an 11-qubit quantum computer. *Nat Commun* **10**, 5464, doi:10.1038/s41467-019-13534-2 (2019).

7 Mirhosseini, M., Sipahigil, A., Kalaee, M. & Painter, O. Superconducting qubit to optical photon transduction. *Nature* **588**, 599-603, doi:10.1038/s41586-020-3038-6 (2020).

8 Higginbotham, A. P. *et al.* Harnessing electro-optic correlations in an efficient mechanical converter. *Nature Physics* **14**, 1038-1042, doi:10.1038/s41567-018-0210-0 (2018).

9 Forsch, M. *et al.* Microwave-to-optics conversion using a mechanical oscillator in its quantum ground state. *Nature Physics* **16**, 69-74, doi:10.1038/s41567-019-0673-7 (2019).

10 Jiang, W. *et al.* Efficient bidirectional piezo-optomechanical transduction between microwave and optical frequency. *Nat Commun* **11**, 1166, doi:10.1038/s41467-020-14863-3 (2020).

11 Rueda, A. *et al.* Efficient microwave to optical photon conversion: an electro-optical realization. *Optica* **3**, doi:10.1364/optica.3.000597 (2016).

12 Fan, L. *et al.* Superconducting cavity electro-optics: a platform for coherent photon conversion between superconducting and photonic circuits. *Science advances* **4**, eaar4994 (2018).

13 Holzgrafe, J. *et al.* Cavity electro-optics in thin-film lithium niobate for efficient microwave-to-optical transduction. *Optica* **7**, doi:10.1364/optica.397513 (2020).

14 Fernandez-Gonzalvo, X., Horvath, S. P., Chen, Y.-H. & Longdell, J. J. Cavity-enhanced Raman heterodyne spectroscopy in Er3+:Y2SiO5 for microwave to optical signal conversion. *Physical Review A* **100**, doi:10.1103/PhysRevA.100.033807 (2019).

15 Bartholomew, J. G. *et al.* On-chip coherent microwave-to-optical transduction mediated by ytterbium in YVO4. *Nat Commun* **11**, 3266, doi:10.1038/s41467-020-16996-x (2020).

16 Hisatomi, R. *et al.* Bidirectional conversion between microwave and light via ferromagnetic magnons. *Physical Review B* **93**, doi:10.1103/PhysRevB.93.174427 (2016).

17 Zhu, N. *et al.* Waveguide cavity optomagnonics for microwave-to-optics conversion. *Optica* **7**, doi:10.1364/optica.397967 (2020).

18 Tsuchimoto, Y. *et al.* Proposal for a quantum interface between photonic and superconducting qubits. *Physical Review B* **96**, doi:10.1103/PhysRevB.96.165312 (2017).

19 Delteil, A. *et al.* Generation of heralded entanglement between distant hole spins. *Nature Physics* **12**, 218-223, doi:10.1038/nphys3605 (2015).

20 Delteil, A., Sun, Z., Falt, S. & Imamoglu, A. Realization of a Cascaded Quantum System: Heralded Absorption of a Single Photon Qubit by a Single-Electron Charged Quantum Dot. *Phys Rev Lett* **118**, 177401, doi:10.1103/PhysRevLett.118.177401 (2017).

21 Wallraff, A. *et al.* Strong coupling of a single photon to a superconducting qubit using circuit quantum electrodynamics. *Nature* **431**, 162-167 (2004).

22 Majer, J. *et al.* Coupling superconducting qubits via a cavity bus. *Nature* **449**, 443-447, doi:10.1038/nature06184 (2007).

23 Landig, A. J. *et al.* Coherent spin-photon coupling using a resonant exchange qubit. *Nature* **560**, 179-184, doi:10.1038/s41586-018-0365-y (2018).

24 Gardiner, C. W. & Collett, M. J. Input and output in damped quantum systems: Quantum stochastic differential equations and the master equation. *Physical Review A* **31**, 3761-3774, doi:10.1103/PhysRevA.31.3761 (1985).

25 Kuhlmann, A. V. *et al.* Transform-limited single photons from a single quantum dot. *Nat Commun* **6**, 8204, doi:10.1038/ncomms9204 (2015).

26 Kuhlmann, A. V. *et al.* Charge noise and spin noise in a semiconductor quantum device. *Nature Physics* **9**, 570-575, doi:10.1038/nphys2688 (2013).

27 Somaschi, N. *et al.* Near-optimal single-photon sources in the solid state. *Nature Photonics* **10**, 340-345, doi:10.1038/nphoton.2016.23 (2016).

28 Langbein, W. *et al.* Radiatively limited dephasing in InAs quantum dots. *Physical Review B* **70**, doi:10.1103/PhysRevB.70.033301 (2004).

29 Samkharadze, N. *et al.* High-Kinetic-Inductance Superconducting Nanowire Resonators for Circuit QED in a Magnetic Field. *Physical Review Applied* **5**, doi:10.1103/PhysRevApplied.5.044004 (2016).
30 Stockklauser, A. *et al.* Strong Coupling Cavity QED with Gate-Defined Double Quantum Dots Enabled by a High Impedance Resonator. *Physical Review X* **7**, doi:10.1103/PhysRevX.7.011030 (2017).
31 Chellu, A. *et al.* GaAs surface passivation for InAs/GaAs quantum dot based nanophotonic devices. *Nanotechnology* **32**, 130001, doi:10.1088/1361-6528/abd0b4 (2021).
32 Tomm, N. *et al.* A bright and fast source of coherent single photons. *Nat Nanotechnol* **16**, 399-403, doi:10.1038/s41565-020-00831-x (2021).
33 Peterer, M. J. *et al.* Coherence and decay of higher energy levels of a superconducting transmon qubit. *Phys Rev Lett* **114**, 010501, doi:10.1103/PhysRevLett.114.010501 (2015).
34 Lecocq, F. *et al.* Control and readout of a superconducting qubit using a photonic link. *Nature* **591**, 575-579, doi:10.1038/s41586-021-03268-x (2021).
35 Youssefi, A. *et al.* A cryogenic electro-optic interconnect for superconducting devices. *Nature Electronics* **4**, 326-332, doi:10.1038/s41928-021-00570-4 (2021).

# Supplementary information:

# Large-bandwidth transduction between an optical single quantum-dot molecule and a superconducting resonator

Yuta Tsuchimoto*, Zhe Sun, Emre Togan, Stefan Fält, Werner Wegscheider, Andreas Wallraff, Klaus Ensslin, Ataç İmamoğlu, and Martin Kroner*

*ETH Zurich, Department of Physics, Zurich, Switzerland*

## Supplementary contents

## I. Device fabrication

The transducer consists of a superconducting coplanar waveguide resonator and single quantum dot molecules (QDM). We detail the fabrication process in this section. First, we grew the heterostructure and self-assembled InAs/GaAs QDMs on a semi-insulating GaAs substrate by molecular beam epitaxy (MBE) in the Stransky-Krastanov mode as shown in Fig. S1a. Step 1 (see Fig. S1b for detailed layer thicknesses). The top (red) and bottom (blue) QDs have different wavelengths so that their electron levels come into a resonance upon applying a bias voltage ($V_{bias}$). The current blocking layer above the QDMs introduces a potential barrier between the top gate and the doped layer and prevents a current from flowing through the device. The bottom DBR mirror (5 pairs of a GaAs/AlAs layer), together with a top Ti/Au gate that we will deposit afterwards, forms a leaky optical cavity. This modifies the QD emission mode such that it couples to a free-space objective lens more efficiently. After characterizing the grown wafer at 4.2 K to check the QD wavelengths and linewidths as well as their density and the cavity wavelength, we cut a piece out from an optimum position on the wafer. Before patterning a superconducting resonator on the chip, we removed all the MBE-grown heterostructures which will overlap with the superconducting circuit. We did this for the following reasons: (i) the doped layer is a significant loss channel for a microwave field. Therefore, we must minimize the capacitive coupling between the doped layer and the superconducting resonator to reduce the loss. (ii) MBE grown layers lying under a superconducting resonator cause irreversible and severe degradation in the *Q* factor of the resonator upon irradiation with light or flowing a current in the device. The reduced *Q* factor can only be recovered by warming the sample up to room temperature. This degradation could be due to charge traps by defects such as EL2 in GaAs[1,2]. The QD and n-doped layers only remain in a mesa (Fig. S1a. Step2) outside the superconducting resonator, and they will be capacitively coupled to the resonator by a small lead. To remove the heterostructures except for the mesa region, we etched the chip by inductively coupled plasma (ICP) etching (Gas: $Cl_2$/Ar/$N_2$) after partially protecting the surface with a photoresist mask. The photoresist mask was carefully removed by hot N-Methyl-2-Pyrrolidone solvent at 80 °C after the etching. Here, we did not use a hard mask and plasma etching to remove the mask because an intense plasma treatment directly on the mesa surface induces spectral fluctuations of QD resonances due to plasma-induced defects. On the mesa structure, we fabricated an ohmic contact to the doped layer by annealing Ge/Au/Ge/Au/Ni/Au at 400 °C (Fig. S1a. Step3). We performed a two-step deposition to pattern a top gate on the mesa and connected it to a superconducting resonator across the edge of the mesa. First, we deposited a 10/150 nm-thick Ti/Au layer over the sidewall of the mesa and the etched GaAs surface, which will act as a lead to the top gate (Fig. S1a. Step4). The thick layer is necessary to connect on the sidewall smoothly since the sidewall of the DBR mirror has large roughness because of the etching rate difference between GaAs and AlAs. Successively, we patterned a 2/15 nm-thick Ti/Au layer (top gate) with the size of ~ 3 μm × 3 μm on the mesa (Fig. S1a. Step 5). We finally patterned a λ/2 superconducting resonator at the vicinity of the mesa by magnetron sputtering of a ~ 50 nm-thick NbTiN film (Fig. S1a Step 6). To maximize the vacuum fluctuation of the microwave electric field at QDs, we positioned the electric field antinode of the resonator at the mesa structure. A bias voltage needed to form a QDM is fed from the bias line located at the field node of the resonator through the central line and the Ti/Au lead (see Fig. 1a in the main text). To block unwanted leakage of microwave photons through this bias line, we implemented a relatively high-impedance line at the field node. Although the architecture of this bias line is suboptimal in terms of attenuation of microwave photons, its leakage is negligible compared with other loss rates (loss at the doped layer and leakage through the external coupling capacitors described in Supplementary Section IV-A) in this device. For better attenuation at the bias line, one can implement a lumped element LC filter[3] or a high-impedance load with a λ/2 line at the field node[4].

## II. Characterization of QDMs

For optical characterization of the QD samples, we perform photoluminescence (PL) and resonant spectroscopy (see Fig. 5 for the schematic of the setup). First, we measured PL maps over the whole gate area (~ 3 μm × 3 μm) by moving the piezo stages and chose two QDMs for the experiments in this paper among numerous QDMs. The excitation wavelength of the laser is 780 nm, and the power is well below the saturation. We detected the PL from the red dots of the QDMs. Fig. S3a and S3b show the peak PL energies as a function of $V_{\text{bias}}$. QDM1 and QDM2 show clear anti-crossings indicating the formation of the molecular states. We fitted the peak energies with the equation

$$\epsilon = \epsilon_{\mathrm{QD}} + \frac{1}{2}\left(\epsilon_{\mathrm{diff}} + e\frac{V_{\mathrm{bias}}}{h_{\mathrm{QD}}}(d_{\mathrm{dir}} + d_{\mathrm{ind}})\right) \pm \frac{1}{2}\sqrt{\left(\epsilon_{\mathrm{diff}} + e\frac{V_{\mathrm{bias}}}{h_{\mathrm{QD}}}(d_{\mathrm{dir}} - d_{\mathrm{ind}})\right)^2 + 4t^2}, \tag{S1}$$

where $\epsilon_{\mathrm{QD}}$ is the inherent energy of a red dot, and the second and third terms describe the energy shifts because of the DC stark shift and the tunnel coupling. The notations of each symbol are the following: $\epsilon_{\mathrm{diff}}$: electron energy difference between the red and blue dots on the lever arm; $e$: elementary charge; $V_{\mathrm{bias}}$: bias voltage; $h_{\mathrm{QD}}$ = 250 nm: distance between the top Ti/Au gate and the doped layer; $d_{\mathrm{dir}}$: dipole size of the direct transition (optical dipole); $d_{\mathrm{ind}}$: dipole size of the indirect transition (permanent dipole); $t$: tunnel coupling between the red and blue dots. $d_{\mathrm{dir}}$ is negligible in this fit because it is two orders of magnitude smaller than $d_{\mathrm{ind}}$. From those fits, we extracted $d_{\mathrm{ind}}$ = 13.0 ± 0.3 nm and 14.3 ± 0.1 nm for QDM1 and QDM2, respectively. These values agree well with the inter dot separation of 12 nm.

To investigate the response under resonant excitation, we performed differential reflection spectroscopy. We scanned the laser frequency around the resonance of the red dots while modulating the $V_{\mathrm{bias}}$ with an amplitude of 0.02 V and a frequency of 3.977 kHz. The laser reflection and the modulated QD emission feed into a photodiode, and a lock-in amplifier demodulates the signal. As shown in Fig. S2c and S2d, we plotted the peak energies as a function of the $V_{\mathrm{bias}}$ and fitted them with Eq. S1. Here, we used the estimated $d_{\mathrm{ind}}$ from the PL as a fixed parameter and others as the initial parameters for the fits. Table S1 summarizes the obtained parameters from the fits. We evaluated the effective dipole sizes from the slopes of the fits and calculated the standard errors by error propagation for the derivative of Eq. S1. The systematic error of $h_{\mathrm{QD}}$ is another source of error for the dipole sizes. However, this does not change the relative agreement between the theoretical line and the experimental points shown in Fig. 2c in the main text. The maximum biases that can be applied, namely, the maximum effective dipole sizes in Figs. S2c and S2d are limited by the charge stability in the QDMs due to tunnel coupling to the doped layer (electron reservoir). One can prevent tunnelling by introducing another current blocking layer between the QDs and the doped layer.

Figure S2e shows the excitonic dipole size $|\boldsymbol{d}|$ dependence of the exciton linewidth $\Gamma_{\mathrm{QD}}^{\mathrm{inh}}/2\pi$ in QDM2, measured by DR spectroscopy. To obtain the $|\boldsymbol{d}|$ dependence, we measured the linewidth at different bias voltages $V_{\mathrm{bias}}$ corresponding to different indirect exciton contents (see the inset). The linewidth gradually increases with increasing $|\boldsymbol{d}|$ because charge noise induces inhomogeneous exciton line broadening via the Stark effect. The contribution from charge noise can be reduced by electric shielding[5] and GaAs surface treatment[6]. There may also be a contribution to the broadening by ionization of the exciton when the bias voltage is close to the edge of the charge stability region[7]. The broadening due to the ionization can be eliminated by choosing an exciton deep inside the charge stability region or introducing another current blocking barrier to prevent electron tunnelling events. We also note that an extremely $|\boldsymbol{d}|$ is not necessary and $|\boldsymbol{d}|$ ~ 2.6 nm is enough to achieve unity internal transduction efficiency with an optimized device, as shown in Figs. 4b and 4c.

## III. Control of the tunnel splitting

As described in the main text, the frequency splitting of an anticrossing $\Delta_{ef}$ should match the resonance frequency $\omega_m$ of the microwave resonator to achieve a high internal conversion efficiency. We controlled $\Delta_{ef}$ by introducing a 6 nm-thick $Al_xGa_{1-x}As$ layer between the QD layers. The total separation of the QDs, including the $Al_xGa_{1-x}As$ and pure GaAs layers, is 13 nm. We changed the composition ratio of $x$ from 0 to 0.20 and measured $\Delta_{ef}$ using PL at 4.2 K. In Fig. S3, we show the minimum $\Delta_{ef}$ obtained at each $x$. The solid red line is an exponential fit. $\Delta_{ef}$ becomes smaller as increasing $x$ and approaches the resolution of our spectrometer. The minimum $\Delta_{ef}/2\pi$ is ~ 30 GHz at $x = 0.20$, suggesting that one can achieve ~ 10 GHz at $x \sim 0.30$. Further investigation at higher $x$ and more statistical analysis are left for future works.

## IV. Characterization of superconducting coplanar waveguide resonators

The superconducting resonator is a typical λ/2 coplanar waveguide resonator. Using an optical microscope, we determined the width of the central line to be $w = 8.0$ μm and the gap between the central line and the ground planes as $s = 7.0$ μm. The resonator length is $l = 3.0$ mm. With the dielectric constant of GaAs at 10 GHz being 10.6[8], we calculated the effective dielectric constant, the geometric inductance, and the capacitance of the resonator as $\varepsilon_{eff} = 5.8$, $L_{gl} = 0.49$ μH/m, and $C_{gl} = 0.13$ nF/m by using conformal mapping techniques[9]. Since NbTiN is a highly disordered material, the kinetic inductance $L_{kl}$ of the resonator is not negligible. By fitting the measured resonance frequency 9.717 GHz at 4.2 K with $\omega_{\mathrm{m}} = 2\pi/2l\sqrt{(L_{\mathrm{kl}} + L_{\mathrm{gl}})C_{\mathrm{gl}}}$, we estimated $L_{\mathrm{kl}}$ = 1.8 μH/m and the characteristic impedance $Z_{\mathrm{c}}$ = 130 Ω. In this calculation, we neglected the coupling capacitances of the finger capacitors $C_{\mathrm{k}}$ and the doped layer $C_{\mathrm{QD}}$ because those capacitances are on the order of $10^{-15}$ F and much smaller than the capacitance of the resonator.

## IV-A. Microwave losses of the device

To investigate the losses of the device, we measured the transmission of several resonators with/without the QD structure by a network analyzer. The resonators were driven at a sufficiently low microwave power not to excite the nonlinearity of the resonators. The transmission was amplified by 60 dB at room temperature to exceed the noise floor of the network analyzer. We calculated $C_{\mathrm{k}}$ of each resonator by using a commercial product (Ansys Maxwell) and the external $Q$ factor ($Q_{\mathrm{ext}}$) by $Q_{\mathrm{ext}} = C_{\mathrm{gl}}l/4\omega_{\mathrm{m}}R_{\mathrm{L}}C_{\mathrm{k}}^2$, where $R_{\mathrm{L}}$ = 50 Ω is an outer load impedance. The internal $Q$ factor ($Q_{\mathrm{int}}$) without the QD mesa structure was estimated to be ~ 1300 at 4.2 K by measuring a resonator with a relatively small $C_{\mathrm{k}}$. If the temperature is low enough, $Q_{\mathrm{int}}$ should be ~10,000, limited by the piezoelectricity of the GaAs substrate[10,11]. When the mesa structure is included in the device and the doped layer couples capacitively to the resonator, $Q_{\mathrm{int}}$ drastically reduces to 86 at 4.2 K. This is because the sheet resistance of the doped layer $R_{\mathrm{s}}$ ~ 200 Ω/sq contributes to the loss through $C_{\mathrm{QD}}$. To confirm this, we simulated the loss of the device using Ansys Electronics Desktop circuit design. We set the bare resonance frequency of 10 GHz, $Z_{\mathrm{c}}$ = 130 Ohm, $C_{\mathrm{k}}$ = 5.2 fF, $C_{\mathrm{QD}}$ = 3.5 fF, and $Q_{\mathrm{int}}$ = 1300 in the absence of the doped layer. $C_{\mathrm{QD}}$ was estimated by simply calculating the capacitance of the two parallel Ti/Au and doped layers. Fig. S4 shows simulated $Q_{\mathrm{L}}$ as a function of $R_{\mathrm{s}}$. For the small $R_{\mathrm{s}}$, $Q_{\mathrm{L}}$ shows a constant value of ~ 600, which is limited by $Q_{\mathrm{ext}}$ and $Q_{\mathrm{int}}$ of the resonator itself. When $R_{\mathrm{s}} \geq 100$ Ω/sq, $Q_{\mathrm{L}}$ drastically reduces, and the loss is dominated by the ohmic loss of the doped layer in this regime. The deviation of the simulated $Q_{\mathrm{L}}$ from the measured value could be stray coupling capacitances between the doped layer and the resonator, which we did not consider in this simulation.

We measured the identical sample again after ~ 2 years. $Q_{\mathrm{int}}$ slightly decreased from 86 to 70, probably due to oxidation of the superconducting film, and the resonance frequency also shifted from 9.717 to 9.676 GHz. In the main text, data taken in both conditions is presented. The total and external decay rates corresponding to $Q_L$ = 86 are $\kappa/2\pi$ = 113 MHz and $\kappa_{ext}/2\pi$ = 8.3 MHz, respectively. The loss rates corresponding to $Q_L$ = 70 are $\kappa/2\pi$ = 138 MHz and $\kappa_{ext}/2\pi$ = 8.3 MHz, respectively. We used these parameters to estimate microwave photon number $n_m$ in the resonator.

Finally, we emphasize that $Q_L$ can be enhanced by replacing the doped layer with a superconductor or a low-loss metal. For instance, one can pick up a mesa structure and transfer it on a pre-patterned back gate on a different substrate by using typical transfer printing techniques[12,13]. We have already fabricated a preliminary device with a transfer printing technique and observed $Q_L \sim 5000$) limited by $Q_{ext}$ and the ohmic loss at a Ti/Au gate (to be published). The loss at the Ti/Au layer can be easily removed by replacing it with a superconductor. $Q_{int}$ should be up to ~ 500,000[14].

## IV-B. Microwave photon number calibration

To estimate the microwave mean photon number ($n_m$) in the resonator, we calibrated the input microwave power $P_m$ on the input port of the sample. In order to calibrate it, we first measured the losses of all the cables and connectors outside of the cryostat by a network analyzer. We estimated the loss of the wires and connectors in the cryostat by measuring the system's transmission with a 50 Ω coplanar waveguide (copper on a printed circuit board) instead of the device. Since the cables and connectors on the input and output lines are identical, we assume that the loss of the input line in the cryostat is simply half of the total loss. Then the estimated loss over the microwave source to the device input port is 10.7 dB at 9.717 GHz.

The microwave photon number inside the resonator was calculated by

$$n_m = \frac{P_m}{\hbar\omega_m}\frac{\kappa_{ext}}{\left(\frac{\kappa}{2}\right)^2 + \Delta_m{}^2}, \tag{S2}$$

where $\Delta_m$ is the frequency detuning between the resonator and the input. We set $\Delta_m = 0$ since the input frequency is the same as the resonator frequency.

## IV-C. Microwave photon number dependence of the resonance

We measured $n_\mathrm{m}$ dependence of the resonance of the superconducting resonator to investigate the tolerance for the strong microwave drive necessary for the sideband and transduction experiments. To mimic the experimental conditions, we illuminated the top gate with a laser (wavelength: ~ 960 nm, power: ~ 50 nW) and applied a bias voltage of $V_\mathrm{bias}$ = 0.30 V while measuring the microwave transmission. We remark that the decrease in $Q_\mathrm{L}$ due to the laser and $V_\mathrm{bias}$ is too small to see, and we have not seen apparent degradation even at the input laser power of 50 uW. Fig. S5 shows the transmission spectra normalized by various $n_\mathrm{m}$. The spectra are identical over the range of $n_\mathrm{m} \leq 2.8 \times 10^7$; thereby, $n_\mathrm{m}$ used in the experiments in the main text does not deteriorate the resonator.

## V. Estimation of the microwave-QDM coupling strength

In Fig. 2b, we showed that sidebands appeared by modulating an exciton with a coherent microwave drive. Fig. S6 displays the extracted amplitudes of the resonance and one of the first sidebands as a function of $n_\mathrm{m}$. Since the stark shift of exciton energy induced by a microwave drive leads to frequency modulation, one can fit the amplitudes by the Bessel function of the first kind $J_l(2g_0\sqrt{n_\mathrm{m}}/\omega_\mathrm{m})$, where $l$ = 0 and 1 correspond to the resonance and its first sidebands. In this fit shown in Fig. S6, the only fitting parameter is the coupling strength $g_0$. The fit captures the oscillation of the amplitudes very well, and the estimated $g_0/2\pi$ = 1.56 ± 0.02 MHz, where the error indicates the standard error of the fit. We performed the same analyses for different dipole sizes and plotted the results in Fig. 2c. To verify that the estimated $g_0$ are reasonable, we compare them with theoretically calculated $g_0$. The single-photon coupling strength is theoretically given by

$$g_0 = \frac{|d|}{\hbar}\sqrt{\frac{\varepsilon_\mathrm{eff}}{\varepsilon_\mathrm{GaAs}}\frac{h_\mathrm{res}}{h_\mathrm{QD}}}E_\mathrm{res}, \tag{S3}$$

where $\varepsilon_\mathrm{eff}$ = 5.8 and $\varepsilon_\mathrm{GaAs}$ = 10.6 are the effective dielectric constant of the resonator calculated by conformal mapping techniques[9] and the dielectric constant of GaAs, respectively. $h_\mathrm{res}$ = 7.0 μm is the distance between the central line and the ground plane of the resonator, estimated by an optical microscope. $h_\mathrm{QD}$ = 250 nm is the distance between the top gate and the doped layer, determined by the calibration of the MBE growth. $E_\mathrm{res}$ is the vacuum fluctuation of the resonator and given by $E_{\mathrm{rms}} = \sqrt{Z_{\mathrm{c}}\hbar\omega_{\mathrm{m}}^2/\pi h_{\mathrm{res}}^2}$. $\sqrt{\varepsilon_{\mathrm{eff}}/\varepsilon_{\mathrm{GaAs}}}$ represents dielectric screening of the electric field. $h_{\mathrm{res}}/h_{\mathrm{QD}}$ accounts for the local electric field enhancement because of the field confinement. The primary error sources in this calculation are $h_{\mathrm{res}}$ and $h_{\mathrm{QD}}$, and we assumed an error of ~ ± 0.5 um for $h_{\mathrm{res}}$. We neglected the error for $h_{\mathrm{QD}}$ since it does not affect the relative agreement between the theoretical line and the experimental values. The experiment and the theory agree well, as shown in Fig .2c; therefore, we conclude that the estimated $g_0$ from the fits are reasonable.

## VI. Microwave-Optical transduction

### VI-A. Derivation of the theoretical transduction efficiency

This section explains the derivation of the theoretical formula for the transduction efficiency. First, we consider the far-detuned case corresponding to the experiment in the main text; the energy detuning between states |e> and |f> ($\Delta_{\mathrm{ef}}$) is much larger than $\omega_{\mathrm{m}}$. Starting with the Hamiltonian Eq.1 in the main text, we write the equations of motion as follows[15,16]:

$$\frac{d}{dt}\sigma_{\mathrm{ge}} = -\left(\frac{\Gamma_{\mathrm{QD}}}{2} + i\omega_{\mathrm{QD}}\right)\sigma_{\mathrm{ge}} - ig_0\sigma_{\mathrm{ge}}\left(b^\dagger + b\right) - \sqrt{\Gamma_{\mathrm{QD}}}\left(\sigma_{\mathrm{gg}} - \sigma_{\mathrm{ee}}\right)a_{\mathrm{L}}, \tag{S4}$$

$$\frac{d}{dt}\sigma_{\mathrm{ge}}b = -\left[\frac{\Gamma_{\mathrm{QD}}}{2} + \frac{\kappa}{2} + i\left(\omega_{\mathrm{QD}} + \omega_{\mathrm{m}}\right)\right]\sigma_{\mathrm{ge}}b - ig_0(n_{\mathrm{m}} + 1)\sigma_{\mathrm{ge}} - \sqrt{\Gamma_{\mathrm{QD}}}\left(\sigma_{\mathrm{gg}} - \sigma_{\mathrm{ee}}\right)ba_{\mathrm{L}}, \tag{S5}$$

$$\frac{d}{dt}\sigma_{\mathrm{ge}}b^\dagger = -\left[\frac{\Gamma_{\mathrm{QD}}}{2} + \frac{\kappa}{2} + i\left(\omega_{\mathrm{QD}} - \omega_{\mathrm{m}}\right)\right]\sigma_{\mathrm{ge}}b^\dagger - ig_0 n_{\mathrm{m}}\sigma_{\mathrm{ge}} - \sqrt{\Gamma_{\mathrm{QD}}}\left(\sigma_{\mathrm{gg}} - \sigma_{\mathrm{ee}}\right)b^\dagger a_{\mathrm{L}}, \tag{S6}$$

where $\sigma_{\mathrm{ge}}$ is the lowering operator between |*g*> and |*e*>, $a_{\mathrm{L}}$ and *b* are the annihilation operators for the laser input field and a microwave photon in the resonator. $\sigma_{\mathrm{gg}}$ and $\sigma_{\mathrm{ee}}$ are the projection operator to state |*g*> and |*e*>. We apply Fourier transformation to the equations in order to move from time to frequency domain. If we consider the response around the red-(blue-)detuned laser frequency, $\sigma_{\mathrm{ge}}b$ ($\sigma_{\mathrm{ge}}b^\dagger$) in Eq. S4 and the last terms of Eq. S5 and S6 are negligible. Then Eq. S5 and S6 read

$$\sigma_{\mathrm{ge}} b(\omega) = -\frac{i g_0 (n_{\mathrm{m}} + 1)}{\frac{\Gamma_{\mathrm{QD}}}{2} + \frac{\kappa}{2} - i(\omega - \omega_{\mathrm{QD}} - \omega_{\mathrm{m}})} \sigma_{\mathrm{ge}}(\omega), \tag{S7}$$

for $\omega \approx \omega^{+} = \omega_{\mathrm{QD}} + \omega_{\mathrm{m}}$ and

$$\sigma_{\mathrm{ge}} b^{\dagger}(\omega) = -\frac{i g_0 n_{\mathrm{m}}}{\frac{\Gamma_{\mathrm{QD}}}{2} + \frac{\kappa}{2} - i(\omega - \omega_{\mathrm{QD}} + \omega_{\mathrm{m}})} \sigma_{\mathrm{ge}}(\omega), \tag{S8}$$

for $\omega \approx \omega^{-} = \omega_{\mathrm{QD}} - \omega_{\mathrm{m}}$. The above equations describe the transition between |$e$, n> and |$e$, n+1> or |$e$, n-1> upon absorbing or emitting a microwave photon. Substituting Eq. S7 or S8 into the Fourier domain of Eq. S4, we obtain

$$\sigma_{\mathrm{ge}}(\omega) = -\frac{2\alpha_{\mathrm{L}}\sqrt{\Gamma_{\mathrm{QD}}}}{\frac{4(n_{\mathrm{m}}+1)g_0^{2}}{\Gamma_{\mathrm{QD}} + \kappa - 2i(\omega - \omega_{\mathrm{QD}} - \omega_{\mathrm{m}})} + \Gamma_{\mathrm{QD}} - 2i(\omega - \omega_{\mathrm{QD}})}, \tag{S9}$$

for the blue-detuned case and

$$\sigma_{\mathrm{ge}}(\omega) = -\frac{2\alpha_{\mathrm{L}}\sqrt{\Gamma_{\mathrm{QD}}}}{\frac{4n_{\mathrm{m}}g_0^{2}}{\Gamma_{\mathrm{QD}} + \kappa - 2i(\omega - \omega_{\mathrm{QD}} + \omega_{\mathrm{m}})} + \Gamma_{\mathrm{QD}} - 2i(\omega - \omega_{\mathrm{QD}})}, \tag{S10}$$

for the red-detuned case. Here we replaced the field operator $a_{\mathrm{L}}$ with an amplitude $\alpha_{\mathrm{L}} e^{-i\omega_{\mathrm{L}} t}$, where $\omega_{\mathrm{L}}$ is the incoming laser frequency and assumed $\sigma_{\mathrm{gg}} - \sigma_{\mathrm{ee}} \approx 1$ (weak excitation approximation). The amplitude of the outcoming field from the QDM is given by $\alpha_{\mathrm{out}} = \alpha_{\mathrm{L}} + \sqrt{\Gamma_{\mathrm{QD}}}\sigma_{\mathrm{ge}}$. One can calculate the internal transduction gain from the absorption efficiency of the incoming blue- or red-detuned photons as $G_{\mathrm{int}} = 1 - (\alpha_{\mathrm{out}}/\alpha_{\mathrm{L}})^{\dagger}(\alpha_{\mathrm{out}}/\alpha_{\mathrm{L}})$, which is equivalent to the heralding emission efficiency. By using the relation $\eta_{\mathrm{int}} = \lim_{n_m \to 0 \text{ or } 1} G_{\mathrm{int}}$, we obtained Eq. 2 in the main text.

Next, we consider the close-to-resonance case ($\Delta_{\mathrm{ef}} \sim \omega_{\mathrm{m}}$). The Hamiltonian of a QDM interacting with a microwave photon and an optical field is given by

$$\begin{aligned} \hat{H} = {} & \hbar\omega_{\mathrm{e}}\hat{\sigma}_{\mathrm{ee}} + \hbar\omega_{\mathrm{f}}\hat{\sigma}_{\mathrm{ff}} + \hbar\omega_{\mathrm{L}}\hat{a}^{\dagger}\hat{a} + \hbar\omega_{\mathrm{m}}\hat{b}^{\dagger}\hat{b} + \hbar g_0(\hat{\sigma}_{\mathrm{fe}}\hat{b}^{\dagger} + \hat{\sigma}_{\mathrm{ef}}\hat{b}) \\ & + \hbar\int d\omega\, \kappa_{\mathrm{QD}}(\omega)[\hat{\sigma}_{\mathrm{ge}}\hat{a}^{\dagger} + \hat{\sigma}_{\mathrm{eg}}\hat{a}], \end{aligned} \tag{S11}$$

Here $\omega_e$ and $\omega_f$ represents the frequencies of the excitonic states |e> and |f>, respectively. This Hamiltonian accurately describes the system only within the rotating wave approximation $|\Delta_{ef} - \omega_m| << \Delta_{ef} + \omega_m$. For optical-to-microwave transduction, an incoming optical photon resonates with the |g> → |e> transition. The excited state |e> emits a microwave resonator photon via the fifth term in Eq. $S11$ upon a decay event |f> → |g>. If we assume the microwave resonator is initially in its ground state $n_m = 0$, the corresponding equations of motion are

$$\frac{d}{dt}\sigma_{\mathrm{ge}} = -\left(\frac{\Gamma_{\mathrm{QD}}}{2} + i\omega_{\mathrm{e}}\right)\sigma_{\mathrm{ge}} - ig_0\sigma_{\mathrm{gf}}b - \sqrt{\Gamma_{\mathrm{ge}}}\left(\sigma_{\mathrm{gg}} - \sigma_{\mathrm{ee}}\right)a_{\mathrm{L}}, \tag{S12}$$

$$\frac{d}{dt}\sigma_{\mathrm{gf}}b = -\left[\frac{\Gamma_{\mathrm{QD}}}{2} + \frac{\kappa}{2} + i(\omega_{\mathrm{f}} + \omega_{\mathrm{m}})\right]\sigma_{\mathrm{gf}}b - ig_0\sigma_{\mathrm{ge}}. \tag{S13}$$

Here we assume that the exciton recombination rates from states |e> and |f> are identical. We define $\Delta_{ef} = \omega_e - \omega_f$ and $\delta = \Delta_{ef} - \omega_m$. Under the condition of $\omega_L = \omega_e$, the amplitude of the outcoming field $\alpha_{\mathrm{out}} = \alpha_{\mathrm{L}} + \sqrt{\Gamma_{\mathrm{QD}}}\sigma_{\mathrm{ge}}$ becomes

$$\alpha_{\mathrm{out}} = a_{\mathrm{L}} + \frac{2\alpha_{\mathrm{L}}\Gamma_{\mathrm{QD}}}{\frac{4{g_0}^2}{\Gamma_{\mathrm{QD}} + \kappa - 2i\delta} + \Gamma_{\mathrm{QD}}}. \tag{S14}$$

The internal transduction efficiency is then given by $\eta_{\mathrm{int}} = 1 - (\alpha_{\mathrm{out}}/\alpha_{\mathrm{L}})^{\dagger}(\alpha_{\mathrm{out}}/\alpha_{\mathrm{L}})$. If $\delta = 0$, the efficiency simplifies to $\eta_{\mathrm{int}} = 4C_0/(1+C_0)^2$ where $C_0 = 4{g_0}^2/\left(\Gamma_{\mathrm{QD}} + \kappa\right)\Gamma_{\mathrm{QD}}$.

For microwave-to-optical transduction, an incoming red-detuned optical photon resonates with the |g> → |f> transition. The excited state |f> absorbs a microwave resonator photon via the fifth term in Eq. $S11$ and emits an upconverted optical photon. One can write the equations of motion for $\sigma_{gf}$ and obtain the same equation for the efficiency under the initial conditions of $n_m = 1$ and $\omega_L = \omega_f$.

## VI-B. Optical pump power dependence of the transduction

The number of optical photons ($n_o$) coupled to a QDM should proportionally enhance the transduction efficiency because of the $\chi^{(2)}$ type nonlinearity of an exciton in single QDs (similar to the enhanced gain $G_{int}$ due to the massive number of microwave photons, see Eq. 2). To demonstrate this, we measured the heralding photons (the up- and down-converted photons) at $g_0/2\pi$ = 1.6 MHz (QDM1) as a function of $n_m$ while changing the optical pump power $P_o$. We estimated internal conversion gain $G_{int}$ by the heralding photon rate normalized by the independently measured resonant emission rate without microwave drive. Fig. S7a shows the $P_o$ dependence of $G_{int}$. When $P_o \leq 25$ nW, $G_{int}$ does not change because $P_o$ is in the linear response range of the QDM, which indicates the number of optical photons coupled to the QDM in the cavity is $n_o = 1$. $P_o \geq 250$ nW is not in the linear absorption regime anymore. Thereby, the optical photon absorption rate, which is proportional to the detected resonant emission rate $N_{\omega_{QD}}^{res}$, starts to saturate. In this case, $n_0$ increases as $n_o \propto P_o/\hbar\omega_{QD}N_{\omega_{QD}}^{res}$, enhancing $G_{int}$ because of the improved coupling strength due to $n_o > 1$. For various $P_o$, we estimated $\eta_{int}$ using the definition $\eta_{int} = \lim_{n_m \to 0 \text{ or } 1} G_{int}$ (see the main text) in the range of $G_{int} \propto n_m$. Fig. S7b shows $\eta_{int}$ proportionally increases as a function of $n_o$, suggesting that the $\chi^{(2)}$ type nonlinearity of an exciton in the QDM enhances $\eta_{int}$ by the collective coupling strength of $\sqrt{n_o}g_0$.

## VII. References

1 Kondratenko, S. V. *et al.* Deep level centers and their role in photoconductivity transients of InGaAs/GaAs quantum dot chains. *Journal of Applied Physics* **116**, doi:10.1063/1.4902311 (2014).

2 Manasreh, M. O. & Fischer, D. W. Quenching and recovery characteristics of the EL2 defect in GaAs under monochromatic-light illumination. *Physical Review B* **40**, 11756-11763, doi:10.1103/PhysRevB.40.11756 (1989).

3 Mi, X. *et al.* Circuit quantum electrodynamics architecture for gate-defined quantum dots in silicon. *Applied Physics Letters* **110**, doi:10.1063/1.4974536 (2017).

4 Chen, F., Sirois, A. J., Simmonds, R. W. & Rimberg, A. J. Introduction of a dc bias into a high-Q superconducting microwave cavity. *Applied Physics Letters* **98**, doi:10.1063/1.3573824 (2011).

5 Somaschi, N. *et al.* Near-optimal single-photon sources in the solid state. *Nature Photonics* **10**, 340-345, doi:10.1038/nphoton.2016.23 (2016).

6 Chellu, A. *et al.* GaAs surface passivation for InAs/GaAs quantum dot based nanophotonic devices. *Nanotechnology* **32**, 130001, doi:10.1088/1361-6528/abd0b4 (2021).

7 Kroner, M. *et al.* Voltage-controlled linewidth of excitonic transitions in a single self-assembled quantum dot. *Physica E: Low-dimensional Systems and Nanostructures* **32**, 61-64, doi:10.1016/j.physe.2005.12.096 (2006).

8 Jones, S. & Mao, S. THE DIELECTRIC CONSTANT OF GaAs AT MICROWAVE AND MILLIMETER WAVE FREQUENCIES. *Applied Physics Letters* **11**, 351-353, doi:10.1063/1.1755010 (1967).

9 Göppl, M. *et al.* Coplanar waveguide resonators for circuit quantum electrodynamics. *Journal of Applied Physics* **104**, doi:10.1063/1.3010859 (2008).

10 Toida, H., Nakajima, T. & Komiyama, S. Vacuum Rabi splitting in a semiconductor circuit QED system. *Phys Rev Lett* **110**, 066802, doi:10.1103/PhysRevLett.110.066802 (2013).

11 Frey, T. *et al.* Characterization of a microwave frequency resonator via a nearby quantum dot. *Applied Physics Letters* **98**, doi:10.1063/1.3604784 (2011).

12 Cheng, C.-W. *et al.* Epitaxial lift-off process for gallium arsenide substrate reuse and flexible electronics. *Nature Communications* **4**, doi:10.1038/ncomms2583 (2013).

13 Jung, Y. H. *et al.* Releasable High-Performance GaAs Schottky Diodes for Gigahertz Operation of Flexible Bridge Rectifier. *Advanced Electronic Materials* **5**, doi:10.1002/aelm.201800772 (2018).

14 Samkharadze, N. *et al.* High-Kinetic-Inductance Superconducting Nanowire Resonators for Circuit QED in a Magnetic Field. *Physical Review Applied* **5**, doi:10.1103/PhysRevApplied.5.044004 (2016).

15 Gardiner, C. W. & Collett, M. J. Input and output in damped quantum systems: Quantum stochastic differential equations and the master equation. *Physical Review A* **31**, 3761-3774, doi:10.1103/PhysRevA.31.3761 (1985).

16 van Enk, S. J. Atoms, dipole waves, and strongly focused light beams. *Physical Review A* **69**, doi:10.1103/PhysRevA.69.043813 (2004).

## Supplementary figures and a table

### Fig. S1. Device fabrication scheme

**a**

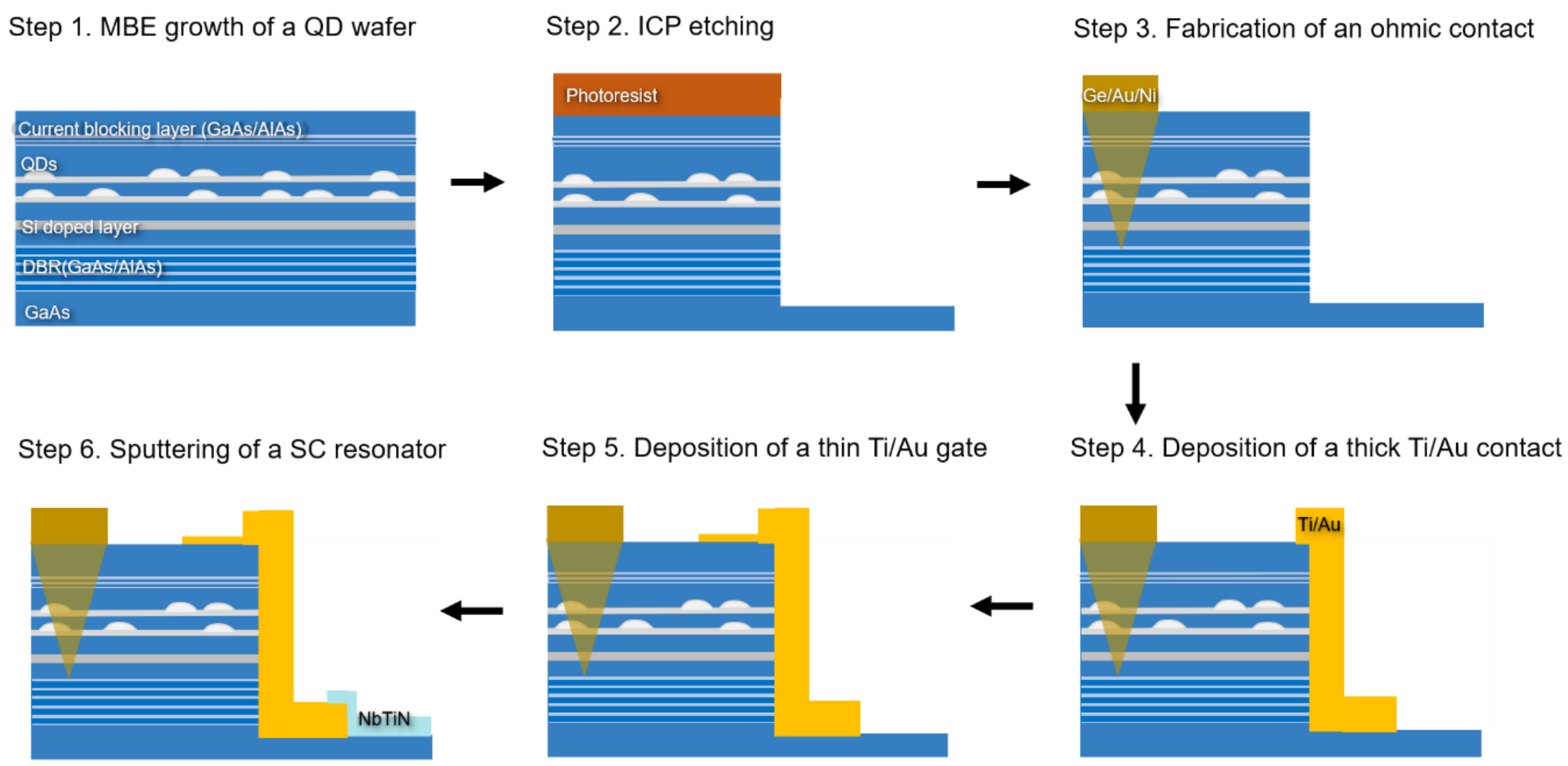

**b**

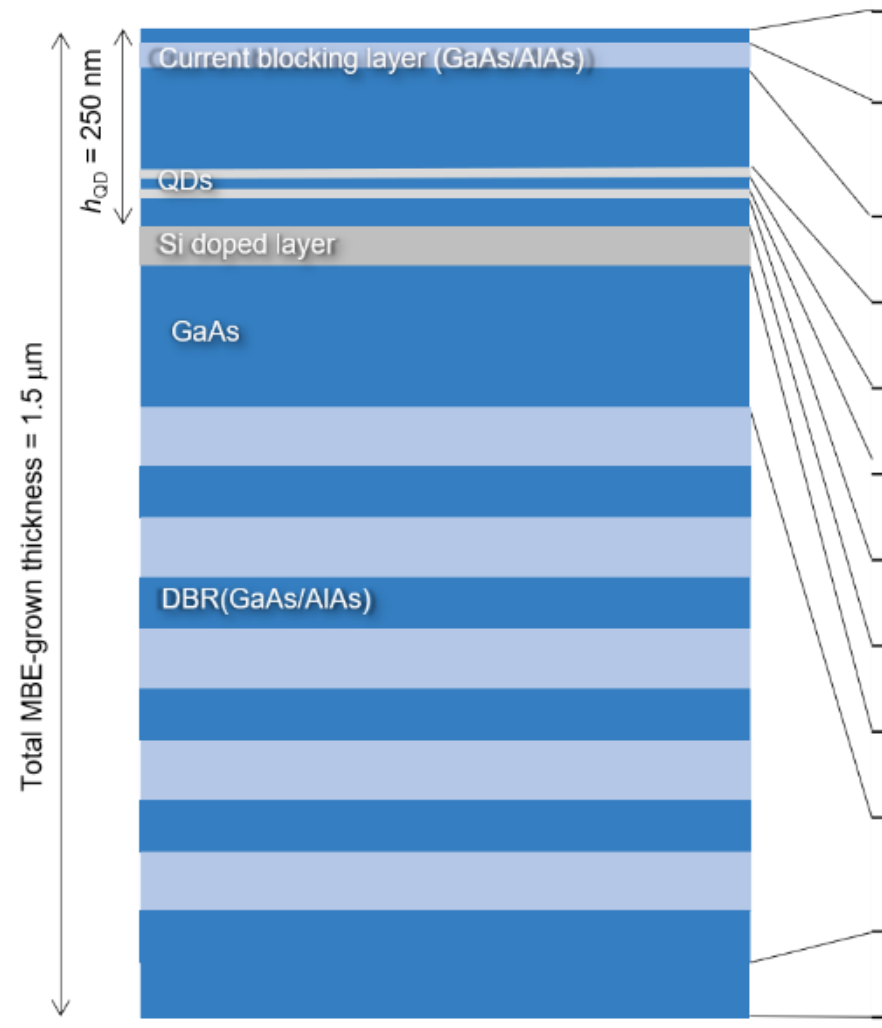

| | Material | Thickness (nm) | |
|---|---|---|---|
| Protection | GaAs | 19 | |
| Current block | GaAs<br>AlAs | 1.9<br>1.9 | × 10 |
| Cap | GaAs | 140 | |
| Top dots | InAs/GaAs | - | |
| Spacer | GaAs | 12 | |
| Bottom dots | InAs/GaAs | - | |
| Tunnel barrier | GaAs | 38 | |
| Back contact | Si:GaAs | 59 | |
| Buffer | GaAs | 200 | |
| DBR | AlAs<br>GaAs | 88<br>73 | × 5 |
| Buffer | GaAs | 190 | |

Fig. S1. **a**, Schematic of the fabrication processes (not to scale). The processes comprise MBE growth of QDMs, ICP etching of unwanted heterostructures, fabrication of ohmic contacts to the doped layer, Ti/Au gate patterning, and patterning of an SC resonator. **b**, Thicknesses of MBE-grown layers. The total thickness is 1.5 μm. The length between the surface and the back contact (doped layer) is $h_{QD}$ = 250 nm.

## Fig. S2. Characterization of QDMs

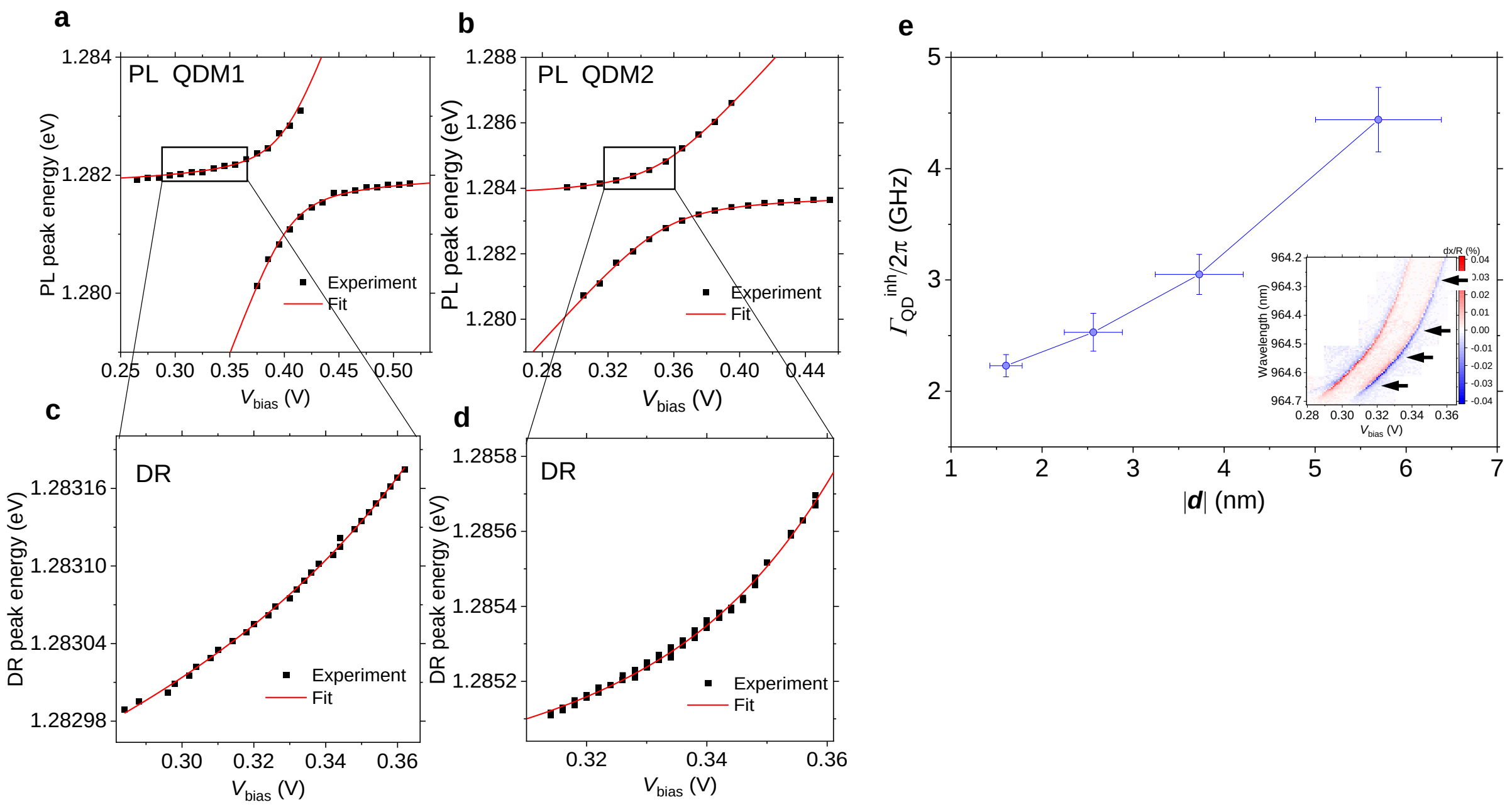

Fig. S2. The top panels show the extracted PL peak energies as a function of bias voltage $V_{\text{bias}}$ for (**a)** QDM1 and (**b)** QDM2. The bottoms are the extracted DR peak energies as a function of $V_{\text{bias}}$ for (**b**) QDM1 and (**d**) QDM2. The black rectangles are the experimental values, and the solid red lines are the fits described in the supplementary text. The energy scales are different between the PL and DR spectra because they were measured with different instruments (grating spectrometer in case of PL and a wavemeter in case of DR) which we have not perfectly calibrated. **e,** dipole size (|$\boldsymbol{d}$|) dependence of the exciton linewidth $\Gamma_{\text{QD}}^{\text{inh}}/2\pi$ of QDM2. The experimentally measured $\Gamma_{\text{QD}}^{\text{inh}}/2\pi$ (blue points) gradually increases against |$\boldsymbol{d}$|. The error bars represent standard errors of the fits used for the estimation of the $\Gamma_{\text{QD}}^{\text{inh}}/2\pi$ and |$\boldsymbol{d}$|. The inset shows a DR map as functions of the laser wavelength and $V_{\text{bias}}$. The black arrows indicate the points where we measured the linewidths.

**Table S1.**

| | $\epsilon_{QD}$ (eV) | $\epsilon_{\mathrm{diff}}$ (meV) | $d_{\mathrm{ind}}$ (nm) | $t$ (meV) |
|---|---|---|---|---|
| QDM1 | $1.2587 \pm 0.0004$ | $24.0 \pm 0.4$ | $13.0 \pm 0.3$ | $1.45 \pm 0.09$ |
| QDM2 | $1.26378 \pm 0.00007$ | $20.89 \pm 0.05$ | $14.3 \pm 0.1$ | $1.01 \pm 0.04$ |

Table S1. Extracted parameters from the fits shown in Fig. S3. The errors are the standard errors of the fits.

**Fig. S3. Control of the tunnel coupling**

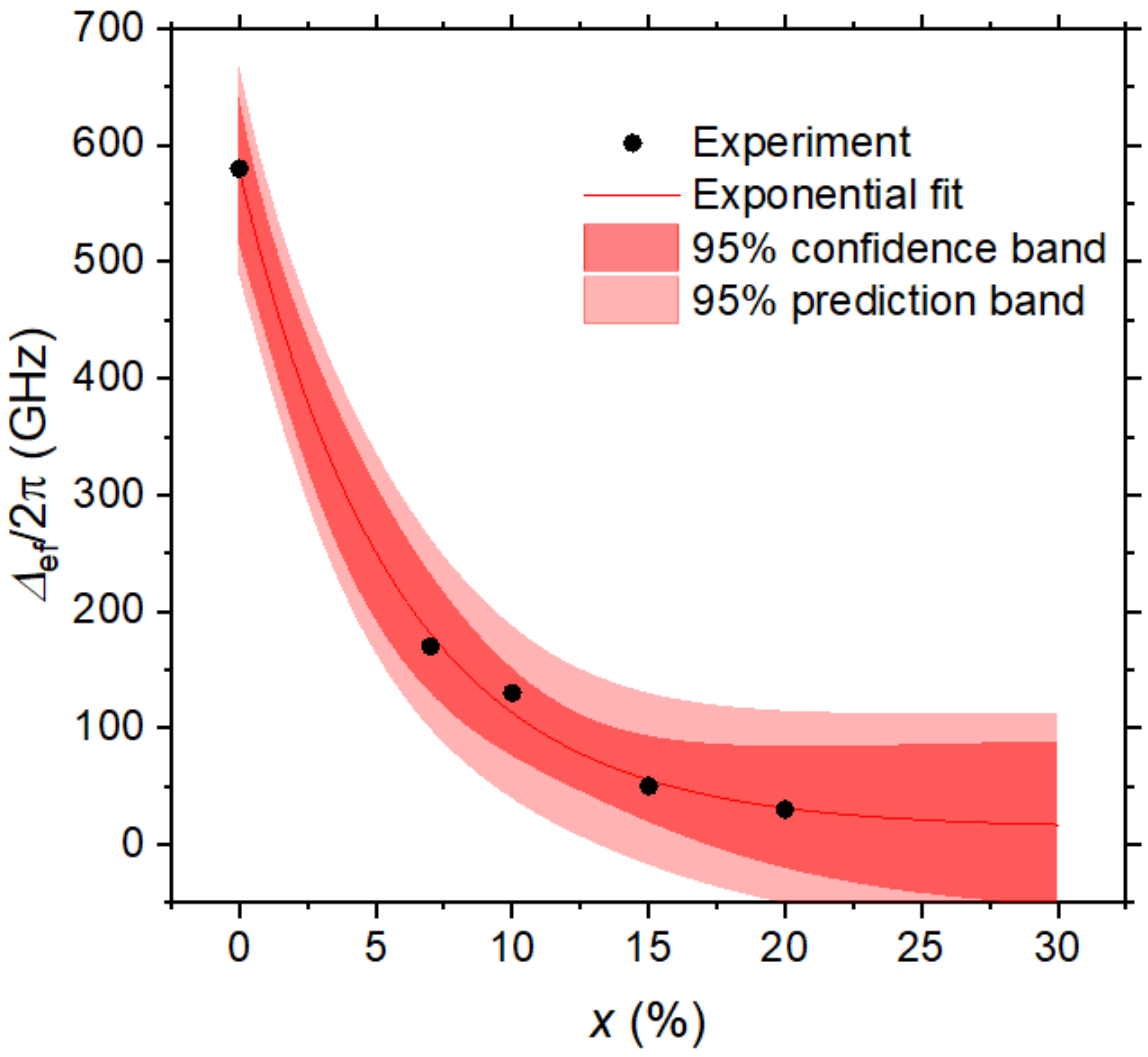

Fig. S3. Experimentally measured tunnel splitting $\Delta_{ef}/2\pi$ as a function of Al composition ratio x in the tunnel barrier layer made of $Al_xGa_{1-x}As$. The black circles are experimental values, and the red line is the exponential fit. The dark and light red areas are the 95% confidence and prediction bands of the fit, respectively.

**Fig. S4. Microwave losses of the resonator**

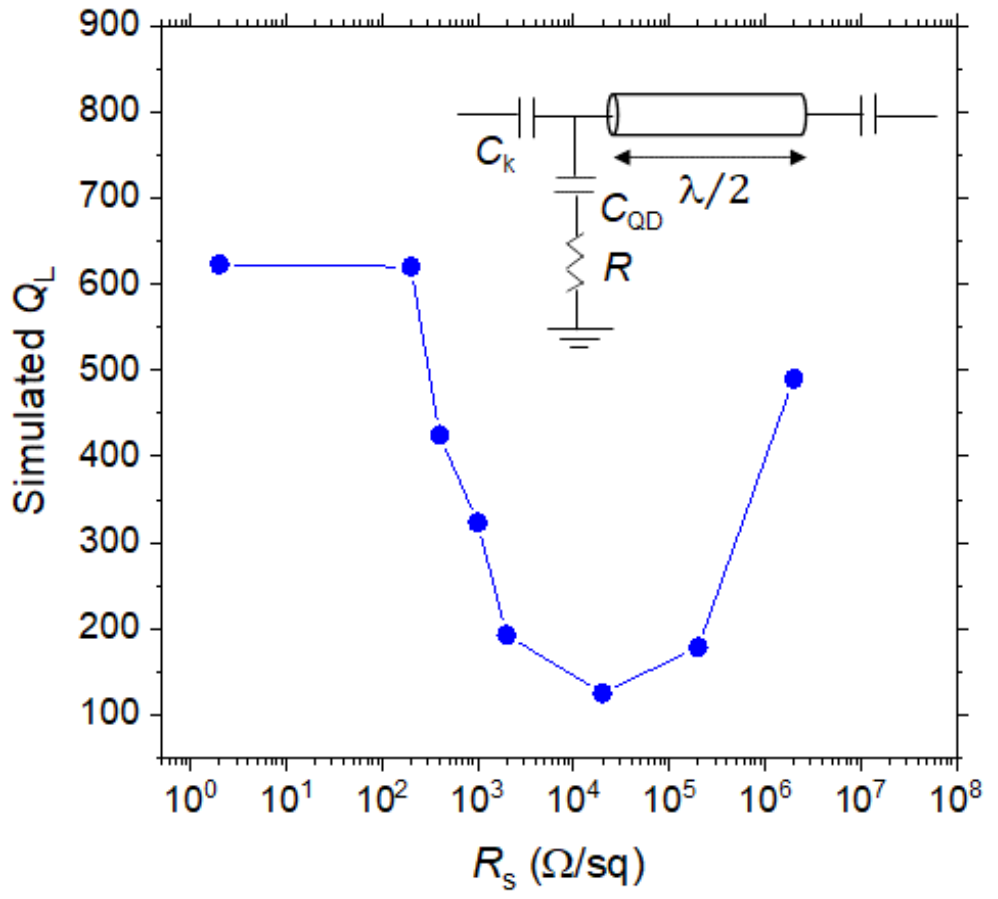

Fig. S4. Simulated $Q_L$ at the various sheet resistances ($R_s$) of the doped layer. The inset shows the circuit used for the simulation.

**Fig. S5. Microwave input power dependence of the resonator**

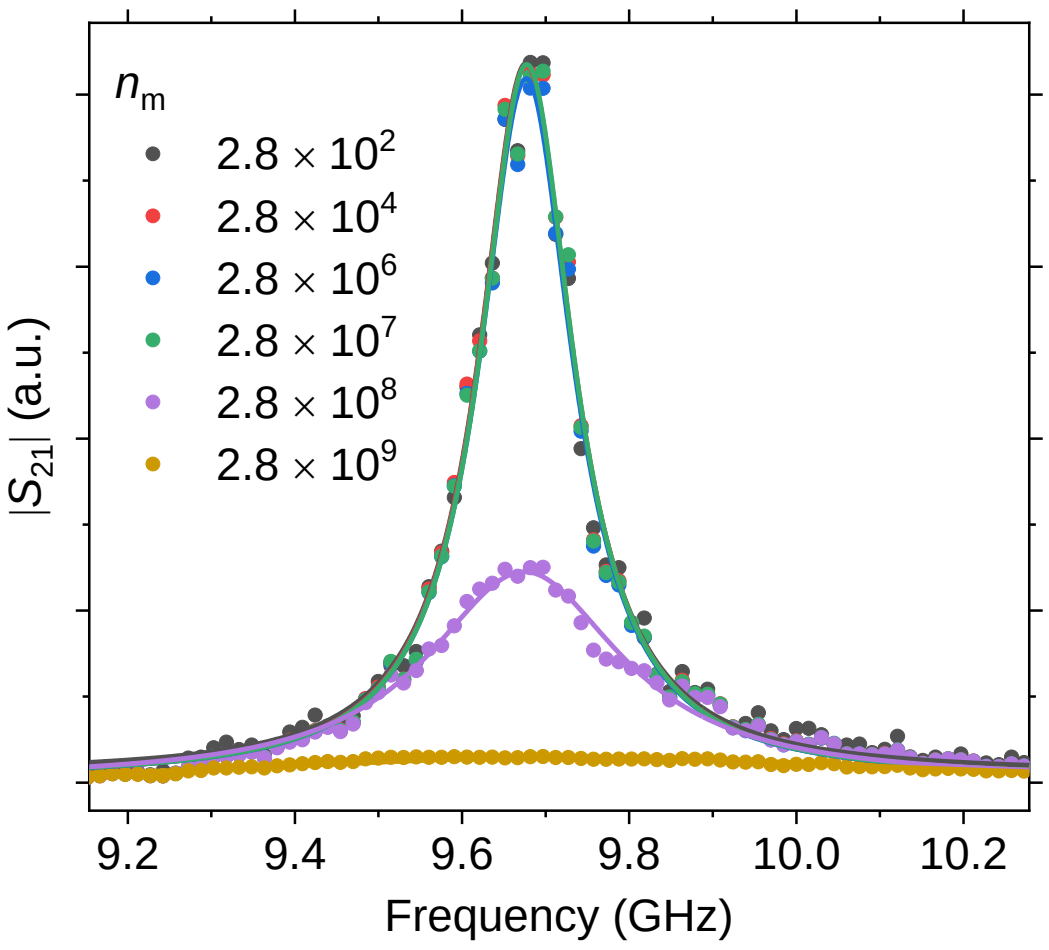

Fig. S5. Microwave mean resonator photon number ($n_m$) dependence of the microwave resonance. Each coloured circle corresponds to the data points at the various input powers. The solid lines are Lorentzian fits for the data.

**Fig. S6. Estimation of QDM-microwave coupling strength**

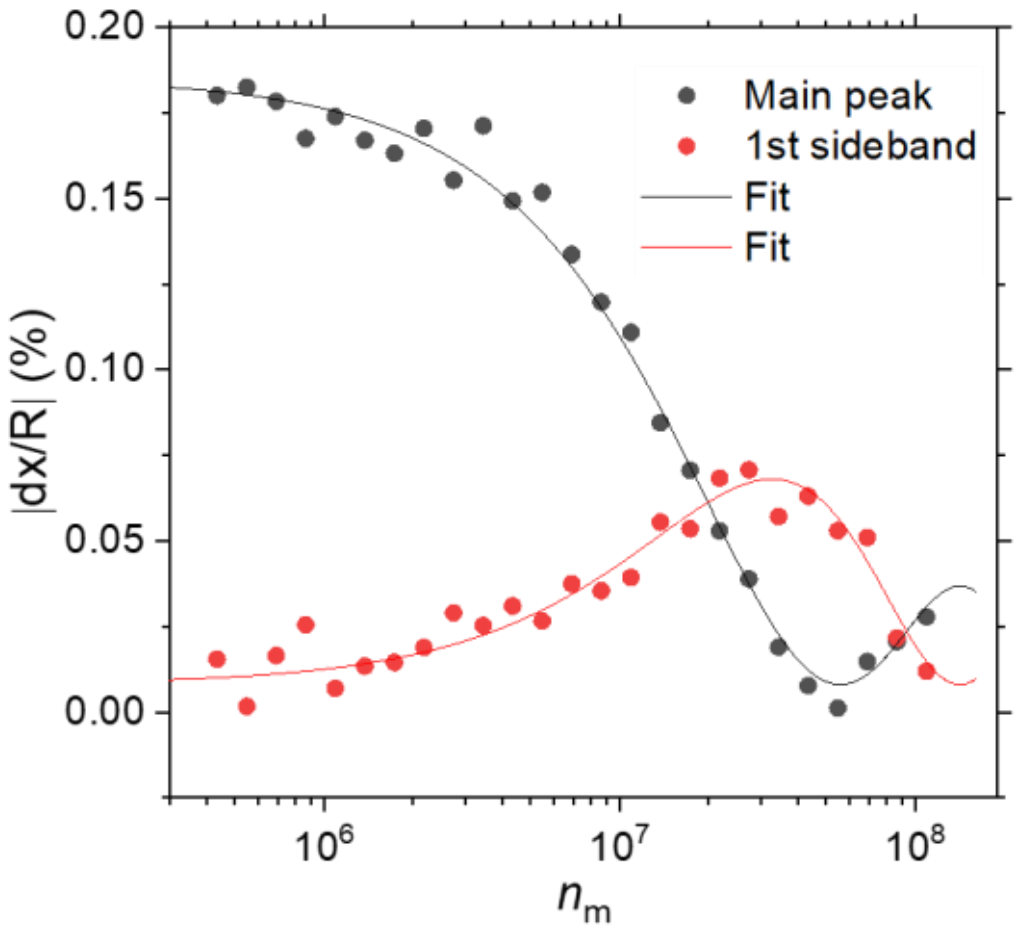

Fig. S6. Microwave mean resonator photon number ($n_m$) dependence of the amplitudes of the main peak (black circles) and its sideband (red circle). The solid lines are the fit with the Bessel function.

**Fig. S7**

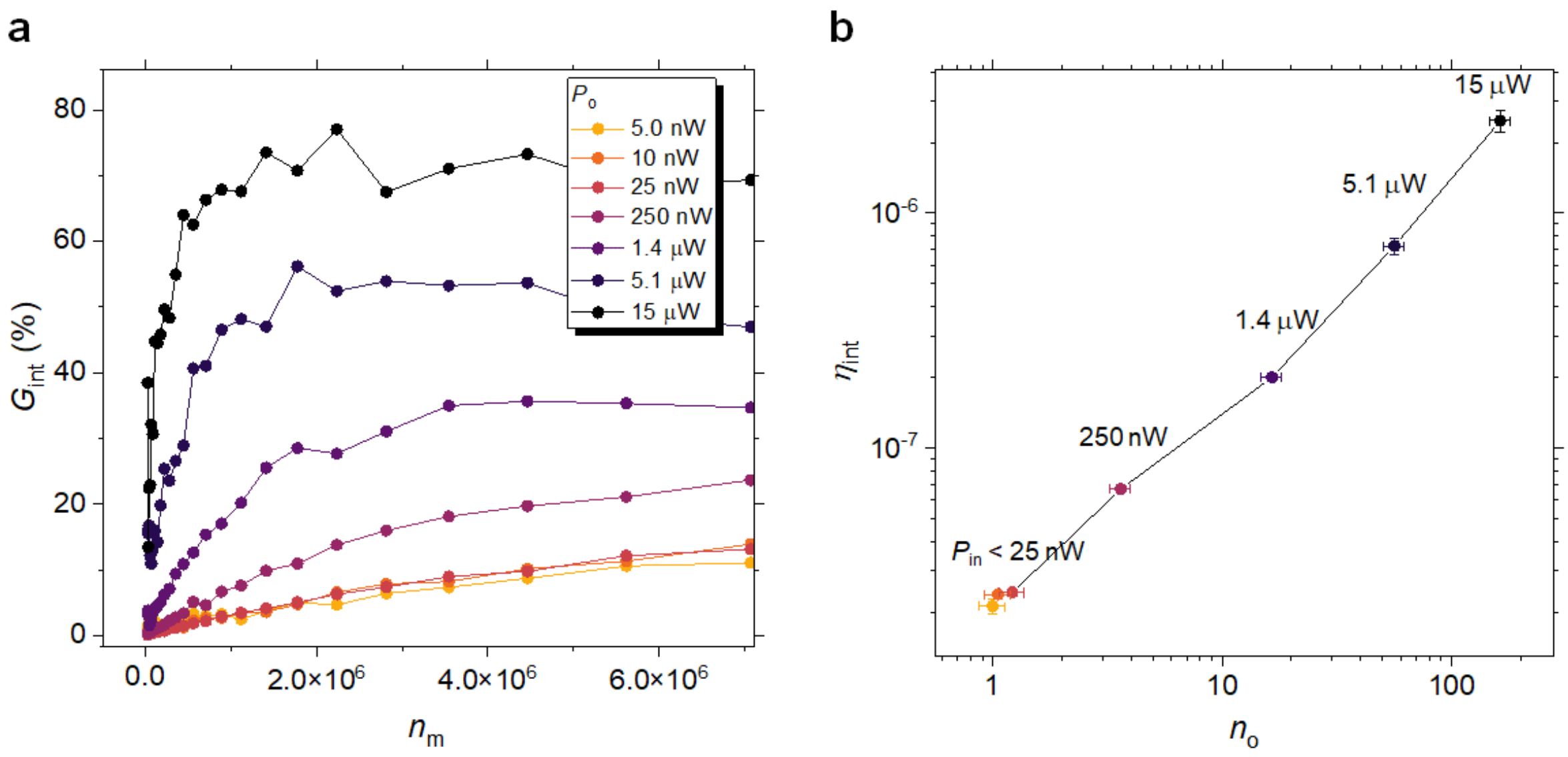

Fig. S7. **a,** Optical power ($P_o$) dependence of internal transduction gain ($G_{int}$) as a function of microwave mean photon number ($n_m$) inside the resonator. Here we used QDM1 at $g_0/2\pi$ = 1.6 MHz. **b**, internal transduction efficiency ($\eta_{int}$) as a function of $n_o$, estimated from **a**.